\documentclass[aps,prl,twocolumn,showpacs,a4paper]{revtex4-1}
\pdfoutput=1

\usepackage{geometry}
\usepackage{color}
\usepackage[latin1]{inputenc}
\usepackage{graphicx}
\usepackage{amsmath,amssymb}
\usepackage{afterpage}

\usepackage{placeins}
\usepackage{mathtools}
\usepackage{textcomp}
\usepackage{amsfonts}
\usepackage{graphicx}
\usepackage{color}
\usepackage[latin1]{inputenc}
\usepackage[svgnames]{xcolor}
\usepackage[
pdftitle={Site-resolved measurement of the spin-correlation function in the Hubbard model},    
colorlinks=True,linkcolor=DarkSlateBlue,citecolor=DarkBlue,urlcolor=DarkBlue,
	pdfstartview=FitH,bookmarks=False,pdfpagemode=UseNone
]{hyperref}

\newcommand\unit[2]{\ensuremath{#1~\mathrm{{#2}}}}
\newcommand\Ket[1]{\ensuremath{|{#1}\rangle}}

\newcommand\Exp[1]{\ensuremath{\langle{#1}\rangle}}
\newcommand\Li{\ensuremath{{^6}\mathrm{Li}}}
\renewcommand{\vec}[1]{\ensuremath{\textbf{#1}}}

\begin{document}
\title{Site-resolved measurement of the spin-correlation function in the Hubbard model}

\date{\today}

\author{Maxwell F. Parsons}
\author{Anton Mazurenko}
\author{Christie S. Chiu}
\author{Geoffrey Ji}
\author{Daniel Greif}
\author{Markus Greiner}
\email{greiner@physics.harvard.edu}
\affiliation{Department of Physics, Harvard University, Cambridge, Massachusetts 02138, USA}

\pacs{
  37.10.Jk,   
  67.85.Lm, 
  71.10.Fd, 
  75.10.Jm, 
  75.78.-n 
}

\begin{abstract}
Exotic phases of matter can emerge from strong correlations in quantum many-body systems.  Quantum gas microscopy affords the opportunity to study these correlations with unprecedented detail.  Here we report site-resolved observations of antiferromagnetic correlations in a two-dimensional, Hubbard-regime optical lattice and demonstrate the ability to measure the spin-correlation function over any distance. We measure the in-situ distributions of the particle density and magnetic correlations, extract thermodynamic quantities from comparisons to theory, and observe statistically significant correlations over three lattice sites.  The temperatures that we reach approach the limits of available numerical simulations.  The direct access to many-body physics at the single-particle level demonstrated by our results will further our understanding of how the interplay of motion and magnetism gives rise to new states of matter.
\end{abstract}

\maketitle 

Quantum many-body systems exhibiting magnetic correlations underlie a wide variety of phenomena.  High-temperature superconductivity, for example, can arise from the correlated motion of holes on an antiferromagnetic (AFM) Mott insulator \cite{Anderson1987a, Lee2006}.  It is possible to probe physical analogs of such systems using ultracold atoms in lattices, which introduce a degree of control that is not available in conventional solid-state systems \cite{Esslinger2010a, Bloch2012}.  Recent experiments exploring the Hubbard model with cold atoms are accessing temperatures where AFM correlations form, but have only observed these correlations via measurements that were averages over inhomogeneous systems \cite{Greif2013a, Hart2014}.  The recent development of site-resolved imaging for fermionic quantum gases \cite{Haller2015a, Cheuk, Parsons2015, Edge2015, Omran2015, Greif2015, Cheuk2016} provides the ability to directly detect the correlations and fluctuations present in a fermionic quantum many-body state.  As demonstrated in experiments with both bosons \cite{Bakr2010, Sherson2010} and fermions \cite{Greif2015, Cocchi2016, Cheuk2016}, microscopy gives access to the spatial variation in observables that occurs in an inhomogeneous system, yielding precise comparisons with theory.  The low energy scales in cold atom systems also allow for time-resolved observations of many-body dynamics, which typically occur on millisecond timescales.  In bosonic systems temporal and spatial resolution have been combined to observe strongly correlated quantum walks \cite{Preiss}, to measure entanglement entropy \cite{Islam2015}, and to study the dynamics of magnetic correlations \cite{Fukuhara2013, Hild2014}.

Here we report the first trap-resolved observations of magnetic correlations in a Fermi-lattice system.  Fermionic atoms in a two-dimensional optical lattice can be well described by the Hubbard Hamiltonian, a simple model in which there is a competition between the nearest-neighbor tunneling energy $t$ and the on-site interaction energy $U$.  Despite the apparent simplicity of the Hubbard model it has a rich phase diagram, containing for example the transition from a metal to a Mott insulator.  AFM spin correlations begin to appear near half-filling when the temperature scale becomes comparable to the exchange energy, which in the strongly interacting regime is $J = 4t^2/U$.  While in three dimensions in the thermodynamic limit there is a finite-temperature phase transition to a state with long-range AFM order, a finite-temperature phase transition is prohibited in two dimensions by the Mermin-Wagner-Hohenberg theorem \cite{Mermin1966}.  Nonetheless, AFM correlations do arise, decaying exponentially with a correlation length $\xi$ that diverges as the temperature goes to zero.  We use quantum gas microscopy to reveal precisely these correlations, which lead to long-range order at a finite temperature where $\xi$ becomes comparable to the finite system size we investigate.

\begin{figure}[th]
\centering
\includegraphics[width=\columnwidth]{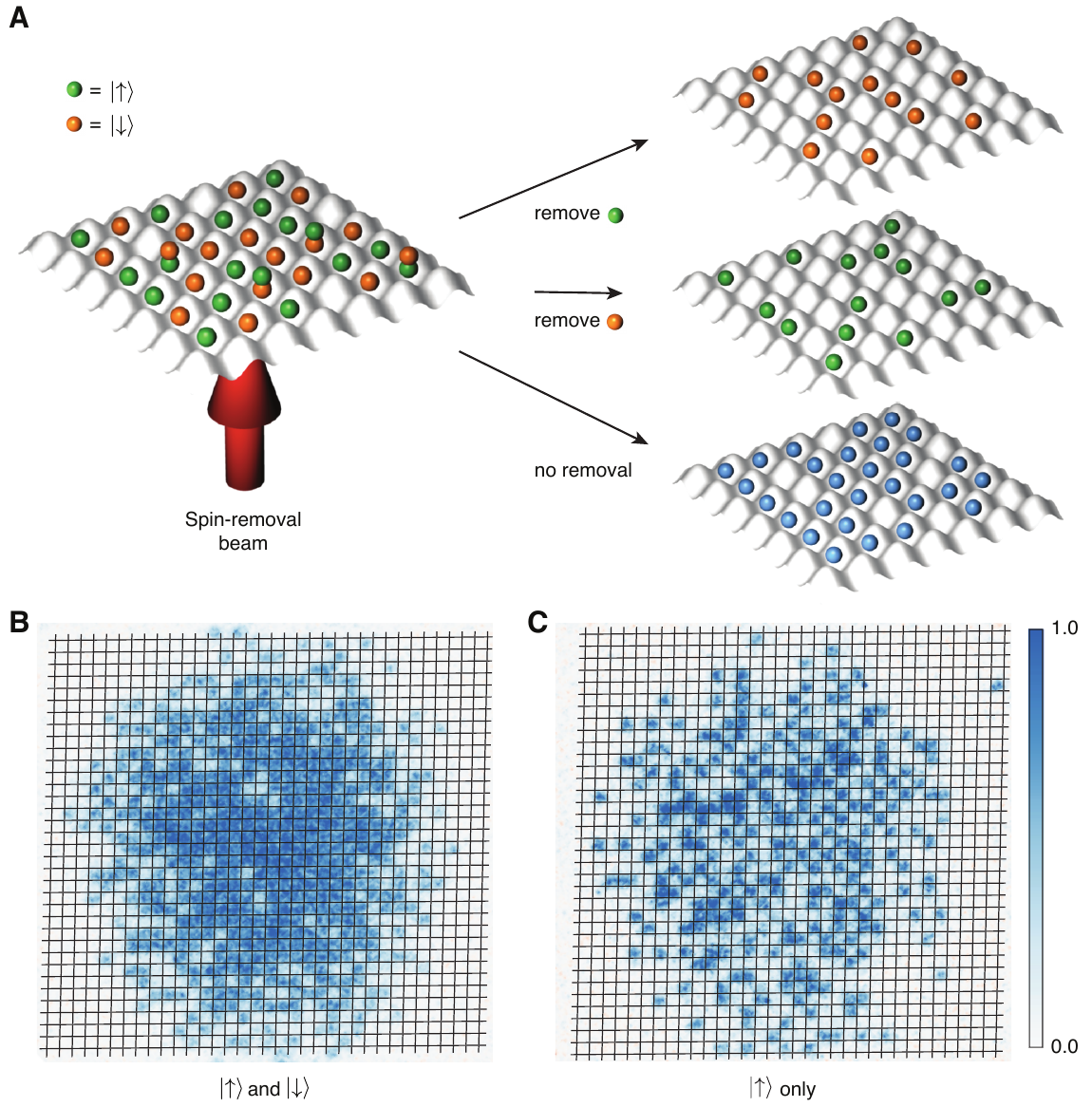}
\caption{\textbf{Experimental technique for measuring spin correlations.}  (A) After loading the atoms into an optical lattice we use a spin-removal technique to map the spin correlations onto charge correlations, which can then be detected using site-resolved imaging.  The two spin states are denoted by green and orange balls.  By driving cycling optical transitions for either spin state with the spin-removal beam we can eject one spin state from the trap.  We can then combine charge correlations measured in images where we remove each spin state and where no removal is performed to compute the local spin correlation\cite{supplementary}. (B) A typical image where no atoms are removed. (C) A typical image with one of the spins removed.
\label{fig:schematic}}
\end{figure}

Our experiments begin with a low-temperature, two-dimensional gas of fermionic $\Li$ atoms in a mixture of the two lowest hyperfine states ($\Ket{\uparrow}$ and $\Ket{\downarrow}$) as described in \cite{Greif2015}.   By adjusting a magnetic bias field in the vicinity of the Feshbach resonance located at \unit{832}{G} we set the s-wave scattering length to $\unit{210}{a_0}$, where $a_0$ is the Bohr radius \cite{Zurn2013}.  Using a \unit{30}{ms} linear ramp, the atoms are adiabatically loaded into an isotropic, square optical lattice with a depth of $s_0 = 7.7(5)\,E_{\mathrm{r}}$, where the recoil energy is $E_{\mathrm{r}}/h = \unit{25.6}{kHz}$ with Planck constant $h$.  We detect magnetic correlations by removing atoms in either spin state and measuring the resulting charge correlations with site-resolved imaging\cite{supplementary}, as shown in Fig. 1.  Because our imaging technique removes doubly occupied sites, both doubly occupied and unoccupied sites show up as empty sites after imaging.  However, proper linear combinations of the different particle and hole correlators (measured both with and without spin dependent removal) will account for the contribution of imperfect unity filling from the signal \cite{supplementary}.  Denoting the observation of a particle (hole) on the site at $\vec{r}$ by $p_{\vec{r}}$ ($h_{\vec{r}}$), we determine the spin correlator\cite{supplementary}
\begin{equation}
\begin{aligned}
C_{\vec{a}}(\vec{r}) &= 4 \Big( \Exp{S^z_{\vec{r}} S^z_{\vec{r}+\vec{a}}} - \Exp{S^z_{\vec{r}}} \Exp{S^z_{\vec{r}+\vec{a}}} \Big) \\
\end{aligned}
\end{equation}
Here, $S_{\vec{r}}^{z} = \frac{1}{2} (n_{\vec{r}}^{\uparrow} - n_{\vec{r}}^{\downarrow})$, with $n_{\vec{r}}^{\sigma}$ denoting the number of particles of spin $\sigma$ on the site at $\vec{r}$.
  We take an average of $C_{\vec{a}}(\vec{r})$ over all $\vec{a}$ where $|\vec{a}| = d$ to obtain $C_{d}(\vec{r})$. The nearest-neighbor, diagonal next-nearest-neighbor, straight next-nearest-neighbor, etc. correlators are given by $C_{1}(\vec{r})$, $C_{1.4}(\vec{r})$, and $C_{2}(\vec{r})$, etc.  From images where neither spin was removed we directly obtain a spatial map of the single occupation probability $\overline{n}_{\mathrm{det}}(\vec{r})$ which also corresponds to the local moment $C_{0}(\vec{r})$.

\begin{figure*}[t]
\centering
\includegraphics[width=\linewidth]{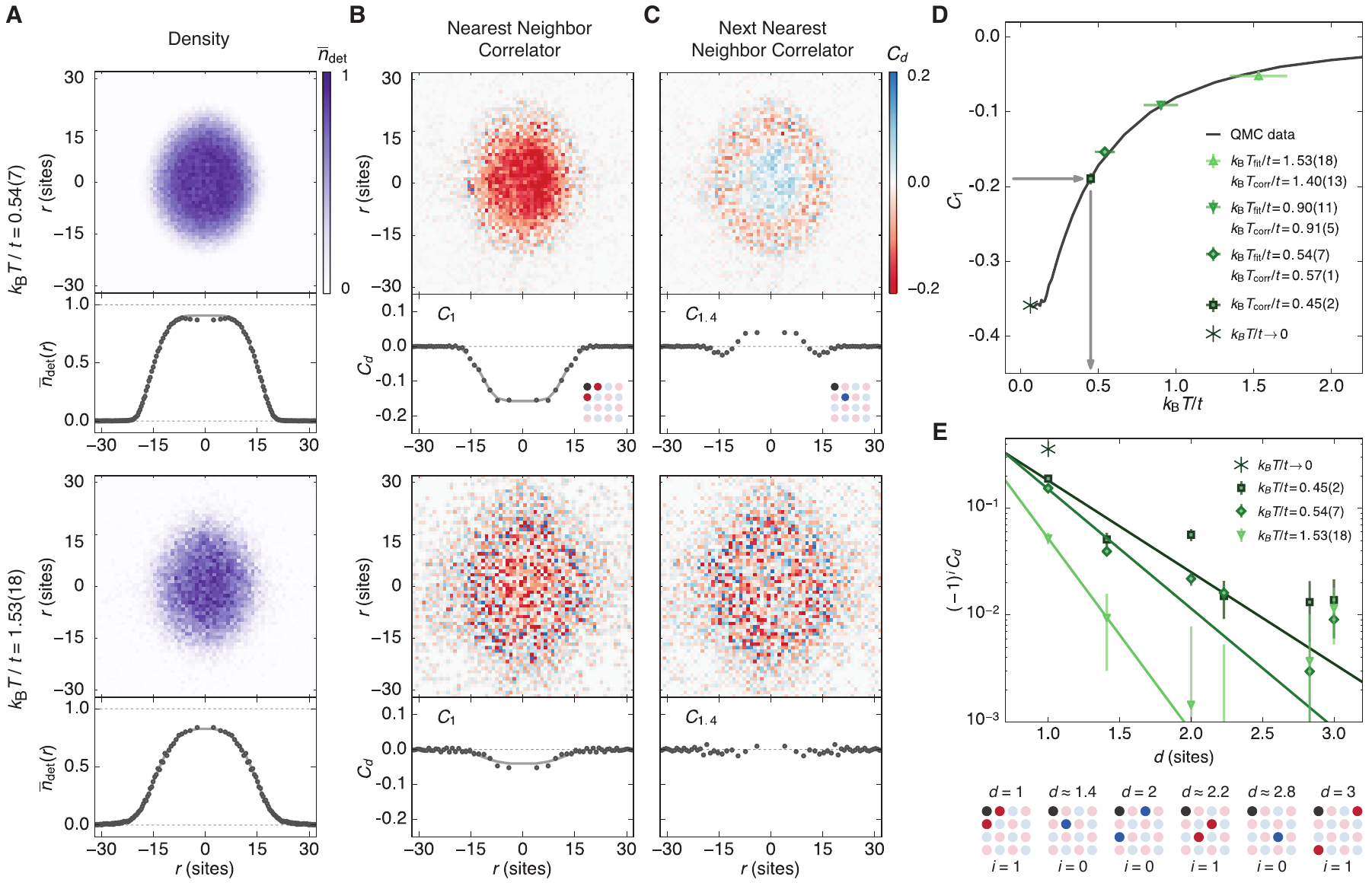}
\caption{\textbf{Local observation of density and spin correlations.} (A-C) We show spatial maps and azimuthally averaged profiles (mirrored about $r=0$ and corrected for ellipticity) of the detected density, nearest-neighbor correlator, and diagonal next-nearest neighbor correlator for a cold (top) and hot (bottom) cloud.  A combined fit, as described in the text, determines the temperature $T$ and chemical potential $\mu$ (solid lines).  (D) We plot the nearest-neighbor correlator in the center of the cloud for samples prepared at different temperatures.  We determine $k_B T_{\mathrm{fit}}/t$ from fits of a numerical linked-cluster expansion to the radial profile and $k_B T_{\mathrm{corr}}/t$ by comparing the central correlator value to a quantum Monte Carlo calculation at half-filling (solid line) \cite{Paiva2010}.  For the coldest data in panels (D) and (E) the NLCE theory error is too large for a fit and we report only the QMC result. (E) An exponential fit to the correlator in the center of the cloud versus $d$ allows us to extract the correlation length for datasets at three different temperatures, giving 0.24(9), 0.39(2), and 0.51(4) sites for decreasing temperature.  The asterisk denotes the nearest-neighbor correlator value from the QMC calculation in (D) as $T \rightarrow 0$.  Error bars on $\overline{n}_{\mathrm{det}}(\vec{r})$ and $C_{d}(r)$ are standard errors after averaging at least 20 sets of combined correlation maps and averaging azimuthally\cite{supplementary}.  All data shown are at $U/t = 8.0(1)$.  Horizontal errors in $d$ are fit errors.}
\label{fig:spatialmaps}
\end{figure*}

After loading atoms into the lattice we observe AFM correlations for nearest-neighbors and diagonal next-nearest-neighbors.  These correlations are strongest in the cloud center where the local chemical potential is set to approximately half-filling.  The spatial maps $\overline{n}_{\mathrm{det}}(\vec{r})$, $C_{1}(\vec{r})$, $C_{1.4}(\vec{r})$ for colder (top) and hotter (bottom) temperatures are shown in panels a, b, and c, respectively, of Fig. 2.  For these data the interaction is tuned to $U/h = \unit{6.82(10)}{kHz}$ with $t/h = \unit{850(100)}{Hz}$ ($U/t = 8.0(1)$).  The chemical potential is tuned to approximately $\mu = U/2$ in the center of the cloud for the colder data by adjusting the atom number to maximize $\overline{n}_{\mathrm{det}}$ in the center.  In Fig. 2a the suppression of $\overline{n}_{\mathrm{det}}$ in the center of the cloud is due to the formation of doubly occupied sites and indicates that the chemical potential in the center of the cloud slightly exceeds $U/2$.   To heat the cloud, we hold the atoms in the optical dipole trap for \unit{4}{s} before loading the lattice.  After heating, the maximum detected occupation decreases from 0.89(1) to 0.84(1) with a slight broadening of the density profile, while the largest magnitude of the nearest-neighbor correlator decreases from 0.154(3) to 0.052(6).  In this regime, where the exchange energy is much smaller than both $U$ and the bandwidth, an increase in temperature quickly saturates the entropy available in the spin degree of freedom while creating little entropy in the charge degree of freedom, making the nearest-neighbor correlator much more sensitive than the density to temperature changes.  For the colder data, we observe significant negative correlations in $C_{1.4}(\vec{r})$ away from half-filling, which requires further theoretical investigation.

We take azimuthal averages along the equipotentials of the underlying harmonic trap to obtain $\overline{n}_{\mathrm{det}}(r)$ and $C_{d}(r)$. The resulting profiles are simultaneously fit to the results of a numerical linked-cluster expansion (NCLE) of the 2D Hubbard model under a local density approximation (LDA) \cite{supplementary, Khatami2011, Chiesa2011}.   From these fits we obtain a temperature $k_{B} T/t = 0.54(7)$ and chemical potential $\mu / U = 0.52(1)$ for the cooler data and $k_{B} T/t = 1.53(18)$ and $\mu / U = 0.33(1)$ for the hotter data.  The excellent agreement with theory provides a strong indication that the local density approximation and the assumption of thermal equilibrium are valid.

By evaporatively cooling further prior to lattice loading, we are able to prepare samples with even larger nearest-neighbor correlations. However, for this data, the NLCE theory error is too large away from half-filling for the fit to converge, owing to the low temperature. Because the averaged correlator in the center may not be at exactly half-filling, by comparing this value for the coldest dataset to a quantum Monte Carlo (QMC) calculation at half-filling \cite{Paiva2010}, we can determine an upper bound on the temperature.  The correlator value of $0.190(8)$ gives $k_{B} T/t < 0.45(2)$, the lowest temperature reported for a Hubbard-regime cold atom system.  The QMC calculation also predicts that the nearest-neighbor correlator settles as $T \rightarrow 0$ to a value of $-0.36$.  Our largest measured nearest-neighbor correlation is therefore $53\%$ of the largest predicted value. In Fig. 2d we plot our largest measured value of the correlator for samples prepared at different temperatures, the temperature derived from the NLCE fits where they converge (x-axis), and the QMC upper bound.  We find very good agreement between our data and theoretical prediction, which is consistent with half-filling at the cloud center.

We see statistically significant antiferromagnetic correlations to distances of three sites and the sign of every measured correlator value is consistent with antiferromagnetic ordering.  Our ability to measure correlations at all length scales allows us to directly extract the correlation length, as shown in Fig. 2e.  Samples are prepared at three different temperatures with the atom number optimized to achieve half-filling in the center of the cloud.  Values for the correlator are taken by averaging the spatial maps over a region in the center of the cloud with a 6-site radius.  To determine the correlation length we perform an exponential fit of $(-1)^i C_{d}$ in the center of the cloud versus $d$, where $i=0$ ($1$) if $d$ is such that the two sites are on the same (a different) sublattice. The correlation lengths are $0.24(9)$, $0.39(2)$, and $0.51(4)$ sites for temperatures of $k_B T/t = $ $1.53(18)$, $0.54(7)$, and $0.45(2)$, respectively.   The asterisk shows the QMC prediction of $0.36$ for the nearest-neighbor correlator at half-filling as $T \rightarrow 0$.

Quantum gas microscopy also allows for a detailed study of the thermalization of the atomic cloud when loading into the lattice.  We investigate the formation of spin correlations and the thermalization of the density distribution for different lattice loading times in Fig. 3.  In these data the lattice is ramped on linearly with a varying duration $t_L$.  We determine the radius $r_{\mathrm{max}}$ where $\overline{n}_{\mathrm{det}}(r)$ is maximized.  For a cloud in thermal equilibrium with $r_{\mathrm{max}}$ not in the center of the cloud, $r_{\mathrm{max}}$ corresponds to half-filling ($\mu = U/2$).  Fig. 3a shows $\overline{n}_{\mathrm{det}}(r_{\mathrm{max}})$ and $C_{d}(r_{\mathrm{max}})$ as a function of loading time.  The detected density grows from $0.6$ at very short loading times and settles at about $0.9$ for $t_L = \unit{10}{ms}$.  The loading time required for the density to settle also corresponds to the maximum absolute values for both the nearest-neighbor and next-nearest-neighbor spin-correlators.  The matching timescales suggest that the suppression of magnetic correlations at $t_L < \unit{10}{ms}$ is due to the low initial density and not to exchange dynamics.  The density at short loading times is determined by the confinement of the optical dipole trap preceding the lattice loading \cite{Greif2015}.  For loading times larger than $\unit{10}{ms}$ both $(-1)^iC_d(r_\mathrm{max})$ and $\overline{n}_{\mathrm{det}}(r_{\mathrm{max}})$ decay, consistent with heating.  The faster decay of $(-1)^iC_d(r_\mathrm{max})$ is further indication that the spin correlators are much more sensitive than the density to temperature in this regime of parameters.

\FloatBarrier
\begin{figure}[h]
\centering
\includegraphics[width=\columnwidth]{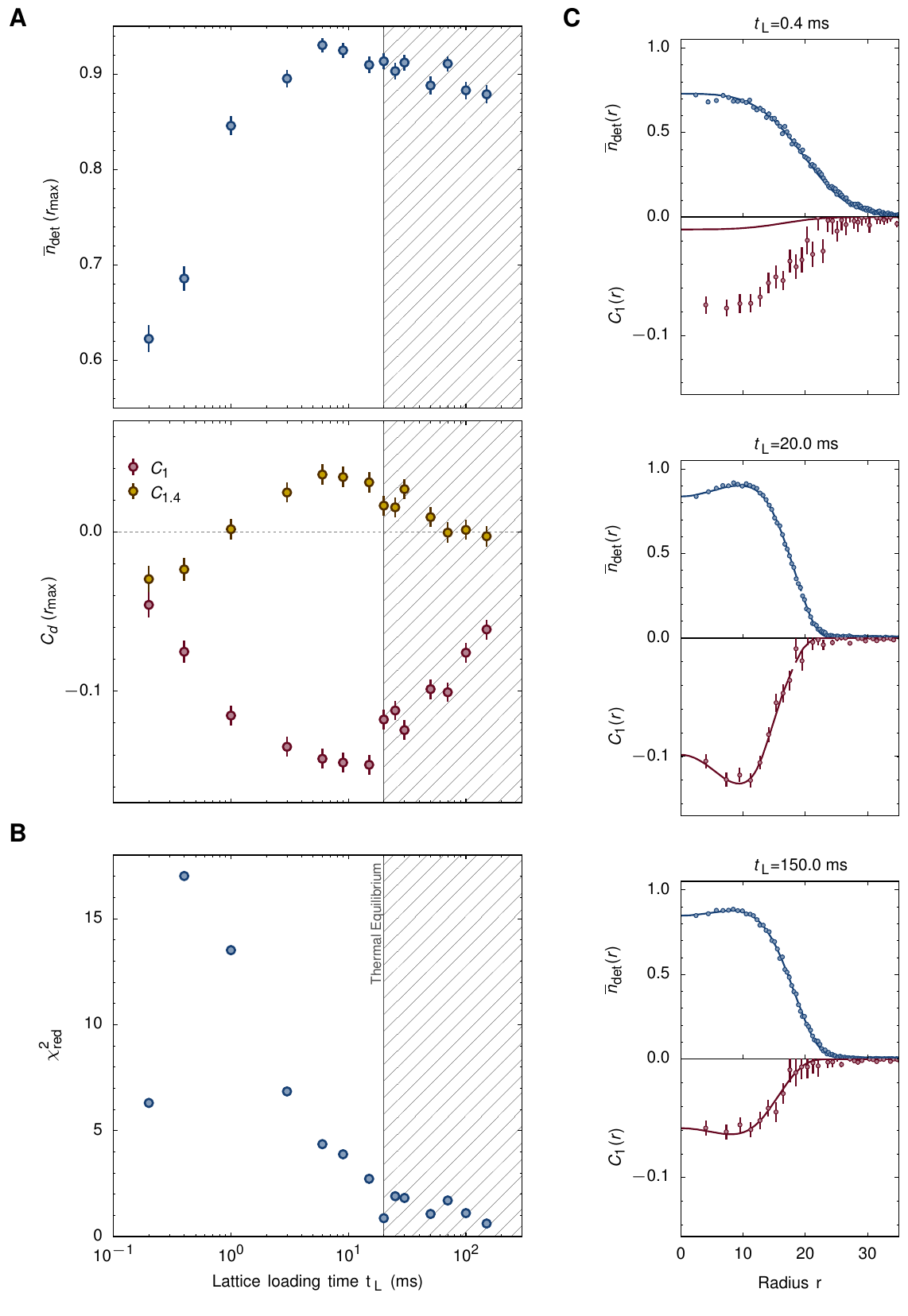}
\caption{\textbf{Thermalization dynamics during lattice loading.} (A) We measure the detected density (upper) as well as the nearest- and next-nearest-neighbor spin correlator at $r_{\mathrm{max}}$ as a function of lattice loading time $t_L$, where $r_{\mathrm{max}}$ is the radius where $\overline{n}_{\mathrm{det}}$ is maximized (lower).  (B) We compute the reduced chi-squared ($\chi^2_{\mathrm{red}}$) of simultaneous fits of the density and nearest-neighbor correlator profiles to NLCE data.  A value $\chi^2_{\mathrm{red}} \approx 1$ indicates a good fit, consistent with our model which assumes thermal equilibrium.  $\chi^2_{\mathrm{red}}$ settles to approximately one at a lattice loading time of \unit{20}{ms}, indicated by the shaded region.  (C) Sample profile fits for three different loading times.}
\label{fig:loadingtime}
\end{figure}

We also study thermalization by fitting the data for different loading times to the NLCE theory and performing a reduced chi-squared ($\chi^2_{\mathrm{red}}$) analysis.  Fig. 3b shows $\chi^2_{\mathrm{red}}$ versus loading time, and Fig. 3c shows individual NLCE fits at $t_L$ of $\unit{0.4}{ms}$, $\unit{20}{ms}$, and $\unit{150}{ms}$ from top to bottom.  The value of $\chi^2_{\mathrm{red}}$ settles to approximately one on a $\unit{20}{ms}$ timescale, which is slightly longer than the settling times for the density and spin correlator.  The value of $\chi_{\mathrm{red}}^{2}$ remains near unity up to our largest loading times, showing that the density and spin correlator distributions remain consistent with thermal equilibrium.

\FloatBarrier
\begin{figure}[!ht]
\centering
\includegraphics[width=\columnwidth]{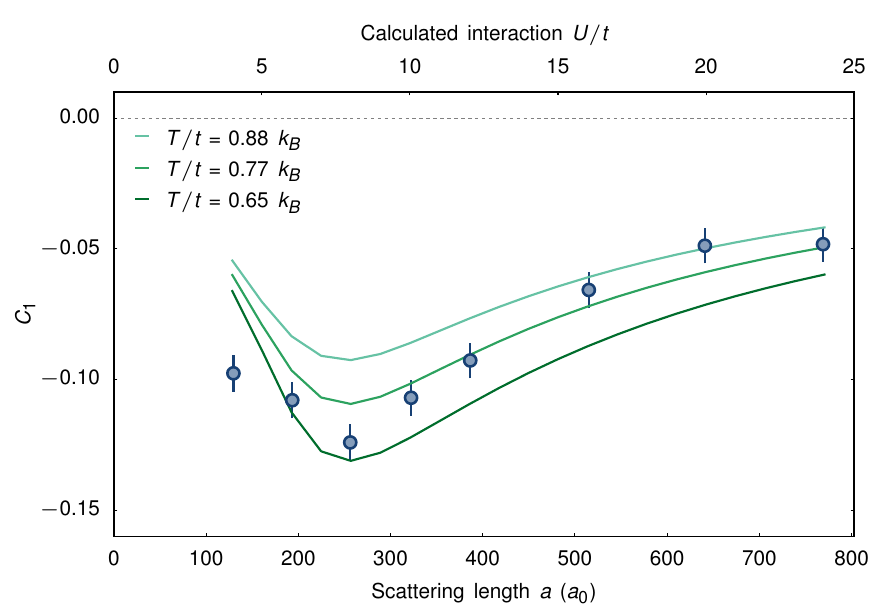}
\caption{\textbf{Varying the interaction strength.} We plot the nearest-neighbor correlator at half-filling for varying scattering length.  The top y-axis gives computed values of $U/t$ for each scattering length\cite{supplementary}.  The solid lines are isothermal theory curves from the NLCE theory.}
\label{fig:interactions}
\end{figure}

While in Bose-Hubbard systems AFM correlations only appear in the Heisenberg limit $U \gg t$, Fermi-Hubbard systems exhibit AFM correlations at all $U/t$, with a maximum in the correlations occurring near $U/t = 8$.  For large $U/t$, AFM correlations are suppressed because the exchange energy becomes small compared to the temperature.  For $U/t < 8$, where the interaction energy is smaller than the bandwidth, charge fluctuations destroy the magnetic correlations.  We study these effects by varying the scattering length for fixed $t/h = \unit{970(110)}{Hz}$.  In Fig. 4 we plot $C_1(r_{\mathrm{max}})$ versus the scattering length, along with the predictions of the NLCE theory for three different temperatures. We show the calculated $U/t$ from Wannier functions in the lowest band, while for our parameters corrections to this single-band approximation could play a role \cite{Buchler2010}. The data shows the expected dependence on $U/t$ from the simple picture mentioned above.  We also compare our data with theoretical isothermal curves at half-filling.  In this comparison additional factors should be considered.  First, the atom number number is fixed, and so the chemical potential in the center of the cloud varies with $U/t$.  Second, we anticipate the loading entropy to be approximately fixed, as opposed to the temperature, and so the data are not expected to strictly follow a single isotherm. The comparison of data with the theory reflects differences between the thermodynamic preparation of atomic and conventional solid-state systems.

Our ability to observe the in-situ, site-resolved distribution of spin correlations at all distances has enabled high-precision comparison with numerical studies and detailed verification that the atomic sample behaves in a manner consistent with thermal equilibrium.  These experimental benchmarks on thermal equilibrium affirm our understanding of the entropy distribution, paving the way for the implementation of entropy redistribution techniques to achieve finite-system-size long-range order \cite{Bernier2009, Ho2009b}. Implementation of such techniques would require precise trap-shaping protocols, which have been fruitfully demonstrated in bosonic quantum gas microscopes \cite{Zupancic2016}.  Numerical simulations provide evidence that a pseudogap phase in the hole-doped 2D Hubbard model arises in conjunction with long-range AFM correlations \cite{Macridin2006} and should therefore be accessible in our experiment in the near future.  At lower temperatures of $T/t \sim 0.03$ a $d$-wave superconductor is expected \cite{Gull2013}.  Further thought is required to understand how the real-space observables that we can measure might shed light on these low-temperature phases.  Beyond equilibrium physics, we could also exploit our ability to take temporally resolved snapshots of the correlations in a many-body wavefunction, allowing for in-depth studies of non-equilibrium physics beyond the capability of existing theoretical tools \cite{Eisert2014}.

\FloatBarrier
\textbf{Acknowledgements} We thank Ehsan Khatami and Marcos Rigol for providing the NLCE calculations, as well as Thereza Paiva and Nandini Trivedi for the Quantum Monto Carlo calculations at half-filling. We also thank J. P. F. LeBlanc and Emanuel Gull for providing additional data based on a dynamical cluster expansion, used for theory verification.  We thank Eugene Demler, Andreas Eberlein, Fabian Grusdt, Jenny Hoffman, Adam Kaufman, M\'{a}rton Kan\'{a}sz-Nagy, Mikhail Lemeshko, and Leticia Tarruell. We thank our colleagues from the Zwierlein group for insightful discussions. Recently, they have also observed antiferromagnetic correlations in their potassium quantum gas microscope (private communication). We acknowledge support from from the Army Research Office Defense Advanced Research Projects Agency Optical Lattice Emulator Program, the Air Force Office of Scientific Research, the Multi University Research Initiative, the Office of Naval Research Defense University Research Instrumentation Program, and NSF. D.G. acknowledges support from the Harvard Quantum Optics Center and the Swiss National Foundation. M.P, A.M. and C.C acknowledge support from the NSF.

\textbf{Author contributions} All authors contributed extensively to the work presented here.

\textbf{Competing financial interests} The authors declare no competing financial interests. 

\textbf{Corresponding authors} Markus Greiner: greiner@physics.harvard.edu

%


\clearpage
\newgeometry{a4paper,
left=0.75in,right=0.75in,top=0.75in,bottom=1.55in
}

\onecolumngrid

\begin{center} 
\begin{Large} \textbf{Methods} \end{Large}
\end{center}
\medskip
\smallskip

\setcounter{section}{0}
\setcounter{subsection}{0}
\setcounter{figure}{0}
\setcounter{equation}{0}
\setcounter{NAT@ctr}{0}

\renewcommand{\figurename}[1]{FIG. }

\makeatletter
\makeatother
\renewcommand{\vec}[1]{{\boldsymbol{#1}}}
\renewcommand\Re{\operatorname{Re}}
\renewcommand\Im{\operatorname{Im}}

\section{Atomic sample preparation}
The procedure used to create the low-temperature Fermi gas and details about the experimental set up can be found in \cite{Greif2015}.

\section{Analysis of site-resolved images}
In place of the kernel deconvolution method that we described in reference \cite{Greif2015}, we are now using a Wiener deconvolution from the scikit-image toolbox \cite{scikit-image} to generate the image from which we determine the site amplitudes for binarization. An example deconvolution, histograms for images with and without spin removal, and a binarized image can be found in Fig. S1.

\section{Spin-removal technique}
After loading the lattice, the lattice depth is ramped linearly to $s_0 = 50\,E_{\mathrm{r}}$ in \unit{1}{ms}, where tunneling is suppressed.  The magnetic bias field is then ramped to \unit{500}{G}.  To remove the atoms in $\Ket{\uparrow}$ ($\Ket{\downarrow}$) a \unit{10}{\mu s} resonant pulse drives a nearly closed transition to the $\Ket{m_J = -3/2, m_I =1}$ ($\Ket{m_J = -3/2, m_I =0}$) state of the $2P_{3/2}$ electronic manifold.  During this pulse $<5 \%$ of the population leaks into the $\Ket{5}$ ($\Ket{4}$) state and remains in the lattice.  A second \unit{10}{\mu s} pulse drives a transition from $\Ket{5}$ ($\Ket{4}$) to $\Ket{m_J = 3/2, m_I =0}$ ($\Ket{m_J = -3/2, m_I =-1}$), ejecting the remaining atoms that were originally in $\Ket{\uparrow}$ ($\Ket{\downarrow}$) from the trap.  Here, we follow the usual convention of labeling the magnetic sublevels of the $2S_{1/2}$ ground electronic manifold $\Ket{1}$ through $\Ket{6}$ in order of increasing energy.

To determine the appropriate pulse durations for removing atoms in $\Ket{\uparrow}$ ($\Ket{\downarrow}$) without removing atoms in  $\Ket{\downarrow}$ ($\Ket{\uparrow}$), we ramp to \unit{500}{G} and apply a resonant pulse as described above for variable time. The resulting loss in atom number follows the sum of two decaying exponentials, corresponding to the loss of the target and non-target spins (see Fig. S2). The characteristic decay times of these exponentials differ by a factor of approximately 1700, reflecting the ratio between resonant and off-resonant scattering rates. We choose the duration of the spin-removal pulse to be equal to ten times the resonant decay time. For this pulse duration, we remove $99.995\%$ of atoms with the target spin while eliminating only $0.6\%$ of the other spin.

\section{Calculation of the spin correlator}
As described in the main text, we measure the polarization-corrected two-point spin correlator, $4(\Exp{S_z S_z} - \Exp{S_z}^2)$, where $S_z = \frac{1}{2}(n_{\uparrow} - n_{\downarrow})$ and lattice site labels for the two-point correlator expectation value entries have been dropped.  Table S1 lists all of our observable two-point correlators using the spin-removal detection scheme, with $d$ denoting a doubly occupied site, $p$ a singly occupied site, and $h$ an unoccupied site.  The expectation value, $\Exp{}_{\mathrm{NR}}$, is over multiple images where neither spin was removed, and the expectation value, $\Exp{}_{R\uparrow}$ ($\Exp{}_{R\downarrow}$), is over multiple images where the atoms in $\Ket{\uparrow}$($\Ket{\downarrow}$) were removed.  In our observed charge correlations we cannot distinguish between unoccupied and doubly occupied sites, so both cases are denoted with $h$ in the formulas below.  By combining correlators in the table, we are able to calculate the polarization-corrected spin correlator in two independent ways:
\begin{equation*}
\begin{aligned}
C'(\textbf{r}) &= \frac{1}{2}\sum_{\sigma\in \{\uparrow, \downarrow\}} \left( \Exp{p p}_{R \sigma} + \Exp{h h}_{R \sigma} - \Exp{p h}_{R \sigma} - \Exp{h p}_{R \sigma}\right) - \Exp{h h}_{NR} - \left(\Exp{p}_{R \uparrow}-\Exp{p}_{R \downarrow} \right)^2\\
C'_{alt}(\textbf{r}) &= 2\sum_{\sigma\in \{\uparrow, \downarrow\}}\Exp{p p}_{R \sigma} - \Exp{p p}_{NR} - \left(\Exp{p}_{R \uparrow}-\Exp{p}_{R \downarrow}\right)^2
\end{aligned}
\end{equation*}
These formulas are incorrect when $\Exp{p}_{R \uparrow} + \Exp{p}_{R \downarrow} \neq \Exp{p}_{NR}$, as in the case of insufficently long or off-resonant removal pulses. The removal of spins is performed in a deep lattice and so is an uncorrelated process. As a result, we can correct for this systematic by removing the uncorrelated contribution to each two-point correlator. For example, the correlator $\Exp{pp}$ is equal to $\Exp{p}^2$ in a sample with no particle correlations. In addition, we remove the polarization correction term $\left(\Exp{p}_{R \uparrow}-\Exp{p}_{R \downarrow}\right)^2$ to avoid double-counting it as it is part of the uncorrelated contribution to the correlators.
\begin{equation*}
\begin{aligned}
C(\textbf{r}) &= &&\frac{1}{2}\sum_{\sigma\in \{\uparrow, \downarrow\}} \Big[ \hspace{-0.5em} && \Exp{p p}_{R \sigma} + \Exp{h h}_{R \sigma} - \Exp{p h}_{R \sigma} - \Exp{h p}_{R \sigma} \\[-1em]
& && && - (\Exp{p}^2_{R \sigma} + (1-\Exp{p}_{R \sigma})^2 - 2\Exp{p}_{R \sigma}(1-\Exp{p}_{R \sigma})) \hspace{0.3em}\Big] \\[-0.5em]
& && \mathrlap{- \left(\Exp{h h}_{NR}-(1-\Exp{p}_{NR})^2\right)}\\
C_{alt}(\textbf{r}) &= && \mathrlap{2 \sum_{\sigma\in \{\uparrow, \downarrow\}} \left(\Exp{p p}_{R \sigma} - \Exp{p}_{R \sigma}^2\right) - \left(\Exp{p p}_{NR}-\Exp{p}_{NR}^2\right)}
\end{aligned}
\label{eq:PP}
\end{equation*}
The following relation between the original and revised correlators holds:
\begin{equation*}
\begin{aligned}
C(\textbf{r}) &= C' + \Exp{h}_{NR}^2 - (1-\Exp{p}_{R\uparrow}-\Exp{p}_{R\downarrow})^2 \\
C_{alt}(\textbf{r}) &= C'_{alt} + \Exp{p}_{NR}^2 - (\Exp{p}_{R\uparrow}+\Exp{p}_{R\downarrow})^2
\end{aligned}
\label{eq:PP}
\end{equation*}
As expected, these additional terms cancel when there is no spin-removal systematic. The effectiveness of these correction factors in removing spin-removal systematics is discussed in a later section.

Calculating the spin-removal and polarization corrections requires estimating the square of expectation values. Unfortunately, directly squaring the mean estimators yields biased estimators for these terms. We can treat measurements of the various densities as independent and identically distributed Bernoulli random variables $X_i, ..., X_n$ with success probability $p$ and $Y_i, ..., Y_n$ with success probability $q$. We calculate the expectation value of the estimators:
\begin{equation*}
\begin{aligned}
\mathbf{E}\left[\widehat{p^2}\right] &= \mathbf{E}\left[\left({1\over n}\sum_{i=1}^n X_i\right)^2\right]\\
&={1\over n^2}\left(\sum_{i \neq j} \mathbf{E}[X_i X_j] + \sum_{i} \mathbf{E}[X_i^2] \right)\\
&=\left(1-{1\over n}\right) p^2 + \left({1\over n}\right) p\\
\end{aligned}
\end{equation*}
\begin{equation*}
\begin{aligned}
\mathbf{E}\left[\widehat{(p \pm q)^2}\right] &= \mathbf{E}\left[\left({1\over n}\sum_{i=1}^n X_i \pm {1\over n}\sum_{i=1}^n Y_i\right)^2\right]\\
&=\mathbf{E}\left[\left({1\over n}\sum_{i=1}^n X_i\right)^2\right] + \mathbf{E}\left[\left({1\over n}\sum_{i=1}^n Y_i\right)^2\right] \pm \mathbf{E}\left[{2\over n^2}\sum_{i, j} X_i Y_j\right]\\
&=\left(1-{1\over n}\right) (p^2 + q^2) + \left({1\over n}\right) (p+q) \pm 2pq\\
&=(p\pm q)^2-{1 \over n}\left(p^2+q^2-p-q\right)\\
\end{aligned}
\end{equation*}
We can then construct bias-free estimators:
\begin{equation*}
\begin{aligned}
\widehat{p^2}' &= \frac{n}{n-1}\widehat{p^2} - \frac{1}{n-1}\hat{p}\\
\widehat{(p\pm q)^2}' &= \widehat{(p\pm q)^2}+{1 \over n} \left(\widehat{p^2}' + \widehat{q^2}' - \hat{p}-\hat{q}\right)
\end{aligned}
\end{equation*}
For our correlator maps, the bias correction for $\widehat{p^2}$ is maximally $0.0125$ and for $\widehat{(p\pm q)^2}$ is $0.025$, which can be significant. When performing azimuthal averages the correction factor is at most $0.0005$.

\section{Data Analysis}
After obtaining the site-resolved population for a set of images with a given spin-removal sequence, we calculate spatial maps of each two-point charge correlation function and the detected on-site density. We average between $20$ and $125$ spatial maps to calculate spatial maps of the average quantities $\Exp{p p}$, $\Exp{p h}$, $\Exp{h p}$, $\Exp{h h}$, and the average single-particle density $\Exp{p}$. We then combine these quantities, as described above, to determine $C(\vec{r})$.  To compute radial profiles of these quantities, we simultaneously average across images and across approximately-equal-area ellipsoidal annuli of similar chemical potential, as described in \cite{Greif2015}. We use annuli containing $35$-$46$ lattice sites for density calculations, and $82$-$92$ sites for spin correlator calculations. During analysis, we scale the coordinate system by the square root of the ratio of the cloud widths along each lattice axis. Reported radii and harmonic trap frequencies are measured in this rotationally-symmetric coordinate system.

\section{Error analysis}
The singles densities are well-described by the mean value of repeated Bernoulli trials, so we use an Agresti-Coull interval to estimate their confidence intervals. Errors for measured densities away from zero and one asymptotically approach the normal approximation to the standard error of the mean. For the number of samples we use, their disagreement is less than $4 \%$ for mean values between $0.01$ and $0.99$, validating our use of the $\chi^2_{\mathrm{red}}$ test statistic to describe goodness-of-fit. For site densities close to zero and one, the error distribution is no longer normal. For this reason, our calculations of $\chi^2_{\mathrm{red}}$ exclude values of the density at radii larger than 18 lattice sites, ensuring that for all datasets the density remains larger than $1\%$.

The individual components of the spin correlator are well-described by the mean value of repeated Bernoulli trials. Since each point in the radial profiles is an average of more than $1000$ data points, we use the central limit theorem to model the error on each component as normally distributed and use standard error propagation techniques to calculate the error on the correlator. For correlator values close to zero, the assumptions of the central limit theorem no longer hold. We show the normal standard error for these points in plots, but apply the same radial cutoff used in the density distributions to calculate the $\chi^2_{\mathrm{red}}$ of our fits.

\section{Spin-Removal Systematics}
We study systematic errors in the spin correlator caused by excessive or insufficient durations of the spin-removal pulse and also by an imperfect imaging fidelity.  To obtain a systematic error on our spin correlator measurements due to fluctuations in the spin-removal efficiency, we measure how the spin correlator varies with spin-removal pulse duration. For fixed interactions and lattice depths, we vary the spin-removal pulse duration between $1.5 \tau$ and $185 \tau$, where $\tau$ is the fast exponential decay time of the spin removal.  We examine the resultant effect on $C$, $C'$, $C_{alt}$, and $C'_{alt}$ for nearest-neighbor correlations in the center of the atomic sample (see Fig. S3). We find that the corrected correlators $C$ and $C_{alt}$ agree for all spin-removal durations and exhibit a relatively small dependence on the spin-removal time. They are maximized around the chosen spin-removal time, and decrease away from that time because information about correlations is lost in the case of incomplete or off-resonant removal. The uncorrected correlators $C'$ and $C'_{alt}$ disagree significantly away from the chosen removal time, because of the systematic on the correlator described in an earlier section. When running the experiment, we measure the exponential decay time of the removal pulse regularly and find that it never varies by more than $20 \%$.  From the measured systematic variation in $C$ with spin-removal duration, we estimate that the fluctuations in our reported correlations due to variation in the spin-removal timescale are at most $0.2 \%$, well within statistical uncertainties.  We also find that the two correlators agree for our chosen spin-removal pulse duration.

In areas of high filling, atoms which hop to neighboring sites during an imaging sequence may create correlated losses. To examine possible systematic errors on the spin correlator due to imperfect imaging fidelity, we vary the number of imaging pulses in our Raman imaging sequence (see \cite{Parsons2015}) and observe the resulting systematic shifts in $C$ and $C_{alt}$ in the center of the cloud (see Fig. S4). The number of pulses is varied between $2.4 \times 10^4$ and $2.56 \times 10^5$, compared to the typical value of $3.2 \times 10^4$ pulses. For these pulse numbers, the imaging fidelity varies linearly between $99.3\%$ and $92.0\%$, respectively. We find a small systematic shift in both correlators of $8.6(4.1) \times 10^-8$ per pulse. Given a typical imaging fidelity of $98 \%$ the fluctuations in the correlator due to imaging are at most $0.04\%$, which is negligible compared to our statistical uncertainties.

\section{Theory comparison}
The system is governed by the Hubbard Hamiltonian, a single band model that includes nearest-neighbor tunneling parameterized by the tunneling energy $t$, and an on-site interaction parametrized by the interaction energy $U$:
\begin{equation}
\begin{aligned}
\hat{H} = & -t\sum_{\Exp{ij}, \sigma} (\hat{c}^{\dagger}_{i\sigma} \hat{c}_{j\sigma} + h.c.) + U\sum_{i} \hat{n}_{i \uparrow} \hat{n}_{i \downarrow} \\
& + \sum_{i, \sigma} (V_i - \mu) \hat{n}_{i\sigma}.
\end{aligned}
\label{HubbardModel}
\end{equation}
Here, $\hat{c}^\dagger_{i\sigma}$ and $\hat{c}_{i\sigma}$ are the fermionic creation and annihilation operators for a particle on site $i$ with spin $\sigma \in \left\{\uparrow, \downarrow\right\}$, $\hat{n}_{i\sigma} = \hat{c}^\dagger_{i\sigma} \hat{c}_{i\sigma}$ is the density operator, $V_i$ denotes the trap energy offset, and $\mu$ is the chemical potential.

To compare our measurements to theory we model our data within a local density approximation (LDA), which states that the entire atomic cloud can be considered as a locally homogeneous system with a spatially varying chemical potential $\mu_\mathrm{LDA}(r)$ as a function of distance $r$ to the cloud center normalized in units of the lattice spacing $l$:
\begin{equation}
\mu_\mathrm{LDA}(r) = \mu - \frac{1}{2} m\overline{\omega}^2 l^2 r^2.
\end{equation}
Here $\mu$ is the chemical potential in the center of the trap, $m$ is the atomic mass of $^6\mathrm{Li}$ and $\overline{\omega}=\sqrt{\omega_x \omega_y}$ is the geometric mean of the trap frequencies along the $x$ and $y$ directions. The confinement of the atomic cloud originates from the underlying Gaussian beam profile of the laser beams forming the optical lattice. As the beam waists in our experiment correspond to approximately $190$ sites and typical cloud radii are about $20$ sites, a harmonic confinement with a quadratic dependence with distance is an excellent approximation. We have verified that including the next order quartic term (which can be estimated from the respective beam waists) does not change any obtained fit parameter throughout the manuscript by more than its individual error bar.

We compare the measured radial profiles of the site occupation $\overline{n}_{\mathrm{det}}(r)$ and nearest-neighbor spin correlator $C_{1}(r)$ to theoretical predictions of the single-band two-dimensional Hubbard model on a square lattice in the grand-canonical ensemble. The theoretical calculation is based on a NLCE, which includes terms up to ninth order that are resummed with Wynn and Euler resummation techniques \cite{Khatami2011}.  The NLCE technique is a comparably powerful in our temperature regime away from half-filling, where quantum Monte Carlo techniques suffer from the sign problem. The NLCE data is given on a dense grid of $(\mu/t, k_{B} T/t, U/t)$ values and we use a linear interpolation between points. For low temperatures and away from half-filling ($\mu=U/2$) deviations in the theoretical predictions are observed between the different resummation techniques (mostly visible as oscillations in the data versus $\mu$).  For all fits and theory profiles shown in the manuscript we omit theory points with an absolute deviation larger than $0.02$ in the occupation and spin correlator value.  For $U/t=8$ this typically restricts reliable fitting to temperatures $k_{B} T/t > 0.5$.

We perform a simultaneous fit of the measured profiles $\overline{n}_{\mathrm{det}}(r)$ and $C_{1}(r)$ to theoretically calculated profiles $\overline{n}^{\mathrm{theory}}_{\mathrm{det}}(r)$ and $C^{\mathrm{theory}}_{1}(r)$, with the temperature $T$ and chemical potential $\mu$ in the center of the harmonic trap as free parameters. The fitting algorithm minimizes the reduced chi-squared value
\begin{equation}
\chi^2_{\mathrm{red}} (T, \mu) = \frac{1}{L} \left[\sum_i \left[ \frac{\overline{n}_{\mathrm{det}}(r_i)-\overline{n}^{\mathrm{theory}}_{\mathrm{det}}(r_i)}{\sigma_{\overline{n}_{\mathrm{det}}(r_i)}}\right]^2 +
\sum_j \left[ \frac{C_{1}(r_j)-C^{\mathrm{theory}}_1(r_j)}{\sigma_{C_1(r_j)}}\right]^2\right].
\end{equation}
Here $L$ is the number of degrees of freedom for the simultaneous fit and $r_i$, $r_j$ denote the radial distances of the individual experimental data points for the site occupation and spin correlation with normal standard errors $\sigma_{\overline{n}_{\mathrm{det}}(r_i)}$ and $\sigma_{C_1(r_j)}$. If an experimental data point coincides with a point where the theoretical prediction becomes unreliable (which occurs for $<10\%$ of the data points in all fits), the reduced chi-squared value is rescaled accordingly. In the main text we quantify the quality of the fit by the reduced chi-squared value at the optimal fit values for parameters $T$ and $\mu$.

\section{System Calibrations}
We calibrate the lattice depth by performing lattice modulation spectroscopy with a non-interacting sample, as described in \cite{Greif2015}. In brief, we determine inter-band resonance frequencies from the spatial widths of the atomic cloud after modulation and subsequent holding time. These resonance frequencies are then compared to band structure calculations to determine the lattice depths along the $x$ and $y$ directions. The nearest-neighbor tunneling matrix elements are calculated from Wannier function overlaps in the single-band approximation. This method is also used to calibrate the lattice depth along the $z$ direction used for creating a single 2D layer.

A calibration of the underlying harmonic trap frequency can be obtained from breathing-mode oscillations, just as in \cite{Greif2015}. After loading a non-interacting gas into the potential of a single-beam lattice, we rapidly decrease the lattice depth and measure the spatial width of the atomic cloud after variable oscillation times. We have verified that additional quartic terms originating from the underlying Gaussian beam profile of the laser beams have a negligible effect on the breathing frequency by varying the excitation amplitude. We repeat this measurement for different lattice depths to obtain a trap frequency calibration. For the lattice depth used in this work we find a geometric mean trap frequency of $\overline{\omega}/2\pi=0.420(30)\,\mathrm{kHz}$, where the error bar is determined from the calibration uncertainty. As explained below, a more precise method based on minimizing the residual error from a theory comparison gives $\overline{\omega}/2\pi=0.439(5)\,\mathrm{kHz}$, which is well within error bars of the calibration. We use the latter value for all fits shown in the main manuscript.

For deep lattices the on-site interaction energy $U$ can be directly calculated from Wannier functions in the lowest band, as the energy gap to the first excited band is much larger than all other Hubbard energy scales. For the lattice depths used in this work of $7.7(5)\,E_{\mathrm{r}}$ ($7.2(4)\,E_{\mathrm{r}}$ for the data in Fig. 4), higher band contributions can play a role and change the value of the on-site interaction energy as compared to the simple lowest-band calculation, which gives $U_0/h=\unit{6.79}{kHz}$ for the standard parameters \cite{Buchler2010}. When comparing our data to theory we find the best agreement (i.e. the smallest $\chi_{\mathrm{red}}^2$ values) for slightly different values of $U$ and $\overline{\omega}$ as compared to the calculated and calibrated values. In Fig. S6 we show for different (fixed) values of $U$ and $\overline{\omega}$ the $\chi_{\mathrm{red}}^2$ obtained from a simultaneous fit to the measured density and spin correlator profiles of Fig. 3 in the main manuscript, with $T$ and $\mu$ as free parameters (see previous section). For an equilibrated gas that can be accurately modeled with a local-density approximation, the exact location of the minimum in $\chi_{\mathrm{red}}^2$ allows a very accurate determination of $U$ and $\overline{\omega}$ for sufficiently precise experimental data with small error bars.  For the fits shown in the manuscript we use $U/h=6.82(10)\,\mathrm{kHz}$ and $\omega / 2\pi=0.439(5)\,\mathrm{kHz}$, which are the parameters with minimal $\chi_{\mathrm{red}}^2$ for the data with the longest lattice loading time of $\tau_{\mathrm{L}}=150\,\mathrm{ms}$. We find very similar values within $5\%$ for the other lattice loading times $\tau_{\mathrm{L}}>20\,\mathrm{ms}$.

\clearpage
\newgeometry{a4paper,
left=0.75in,right=0.75in,top=0.75in,bottom=1.4in
}

\onecolumngrid

\renewcommand{\figurename}[1]{Extended Data Figure } 
\renewcommand{\tablename}[1]{Extended Data Table }

\begin{figure}[h]
	\centering
	\includegraphics{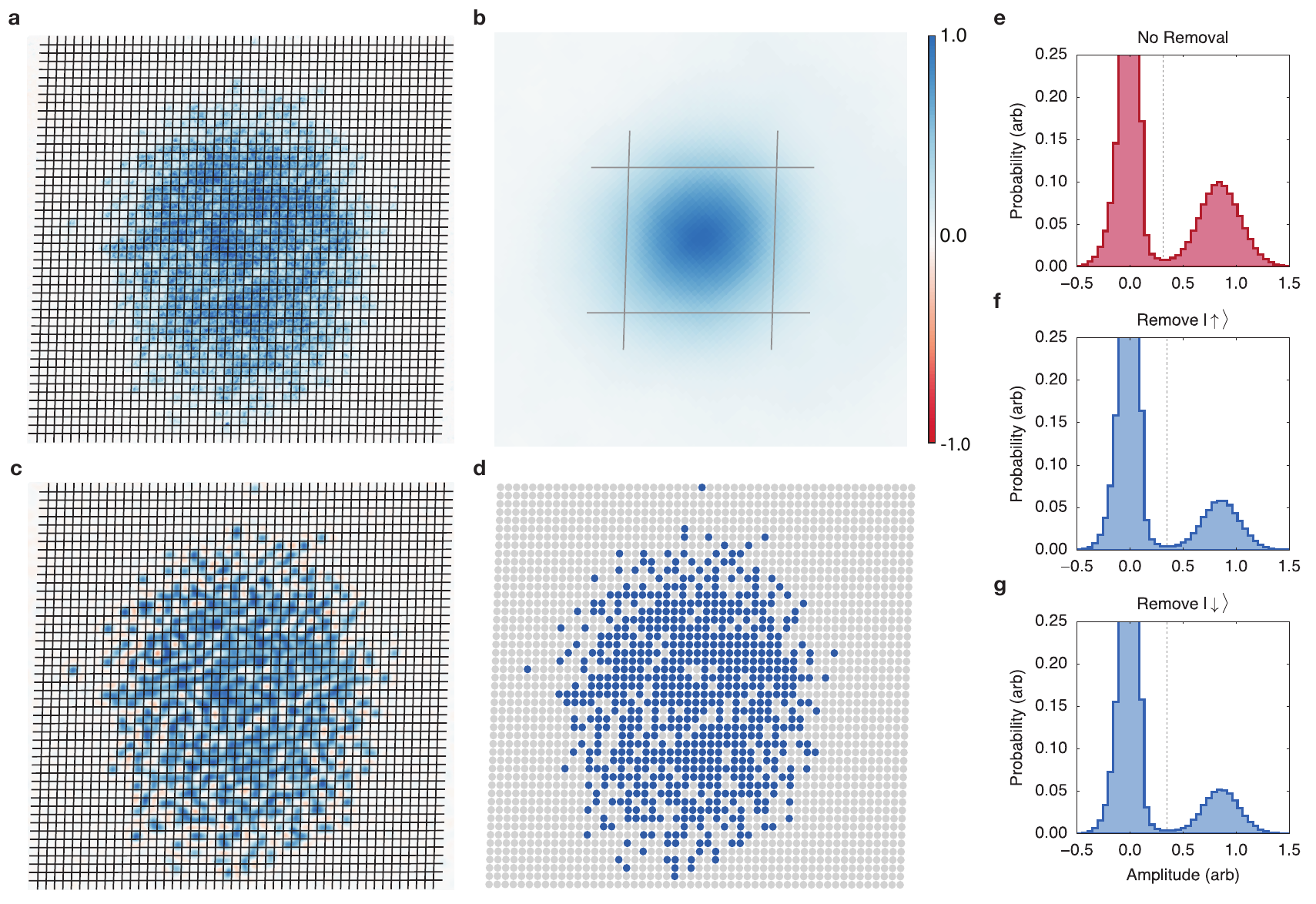}
	\caption{Characterization of Wiener deconvolution for image analysis. (A) Example of a raw image, which is deconvolved with (B), an independently-measured point spread function using the Wiener filter. Grid lines denote the separation of lattice sites, based upon a lattice geometry determined through a Fourier transform (see Analysis of site-resolved images). (C) The deconvolved image, from which lattice site amplitudes can be determined. (D) The resulting binarized image after thresholding the central value at each site based on a histogram of amplitudes. (E, F, G) Histograms of lattice amplitudes for single images in which there was no spin removal, removal of $\Ket{\uparrow}$, or removal of $\Ket{\downarrow}$, respectively. The zero-atom peak is centered around amplitude zero, while the single atom peak is located near amplitude one. As expected, the spin removal process only decreases the height of the single atom peak and has little effect on the threshold value or average amplitude of occupied sites.}
	\label{fig:deconvo}
\end{figure}

\clearpage

\section*{Tab S1}

\begin{table}[h]
\centering
\label{tab:SpinCorrelator}
\begin{tabular}{c c}
\hline
Measured Correlator & Corresponding Spin Correlator\\
\hline
$\Exp{pp}_{NR}$ & $\Exp{\uparrow \uparrow} + \Exp{\downarrow \downarrow} + \Exp{\uparrow \downarrow} + \Exp{\downarrow \uparrow}$\\
$\Exp{ph}_{NR}$ & $\Exp{\uparrow h} + \Exp{\downarrow h} + \Exp{\uparrow d} + \Exp{\downarrow d}$\\
$\Exp{hp}_{NR}$ & $\Exp{h \uparrow} + \Exp{h \downarrow} + \Exp{d \uparrow} + \Exp{d \downarrow}$\\
$\Exp{hh}_{NR}$ & $\Exp{h h} + \Exp{d d} + \Exp{h d} + \Exp{d h}$\\
$\Exp{pp}_{R\uparrow}$ & $\Exp{\downarrow \downarrow}$\\
$\Exp{ph}_{R\uparrow}$ & $\Exp{\downarrow h} + \Exp{\downarrow d} + \Exp{\downarrow \uparrow}$\\
$\Exp{hp}_{R\uparrow}$ & $\Exp{h \downarrow} + \Exp{d \downarrow} + \Exp{\uparrow \downarrow}$\\
$\Exp{hh}_{R\uparrow}$ & $\Exp{\uparrow \uparrow} + \Exp{h \uparrow} + \Exp{d \uparrow} + \Exp{\uparrow h} + \Exp{\uparrow d} + \Exp{h h} + \Exp{d d} + \Exp{d h} + \Exp{h d}$\\
$\Exp{pp}_{R\downarrow}$ & $\Exp{\uparrow \uparrow}$\\
$\Exp{ph}_{R\downarrow}$ & $\Exp{\uparrow h} + \Exp{\uparrow d} + \Exp{\uparrow \downarrow}$\\
$\Exp{hp}_{R\downarrow}$ & $\Exp{h \uparrow} + \Exp{d \uparrow} + \Exp{\downarrow \uparrow}$\\
$\Exp{hh}_{R\downarrow}$ & $\Exp{\downarrow \downarrow} + \Exp{h \downarrow} + \Exp{d \downarrow} + \Exp{\downarrow h} + \Exp{\downarrow d} + \Exp{h h} + \Exp{d d} + \Exp{d h} + \Exp{h d}$\\
\hline
\end{tabular}
\caption[List of observable two-point correlators with the blowout detection method]{List of observable two-point correlators with the spin-removal detection method. This table lists all of the two-point correlators that can be observed with the spin-removal detection method.  Here $p$ refers to particle, $h$ to hole, and $d$ to doublon.  The left column lists the correlators that can be directly observed. Due to parity imaging, we are unable to distinguish between holes and doublons, so the detection of either is denoted $h$. The expectation values $\Exp{}_{NR}$, $\Exp{}_{R\uparrow}$, and $\Exp{}_{R \downarrow}$ denote correlators in images after no spin removal, removing $\Ket{\uparrow}$, and removing $\Ket{\downarrow}$ respectively. The right column shows the two-point spin correlators corresponding to each of the directly measured charge correlators.}
\label{tab:BlowoutCorrelators}
\end{table}

\clearpage

\begin{figure}[h]
\centering
\includegraphics{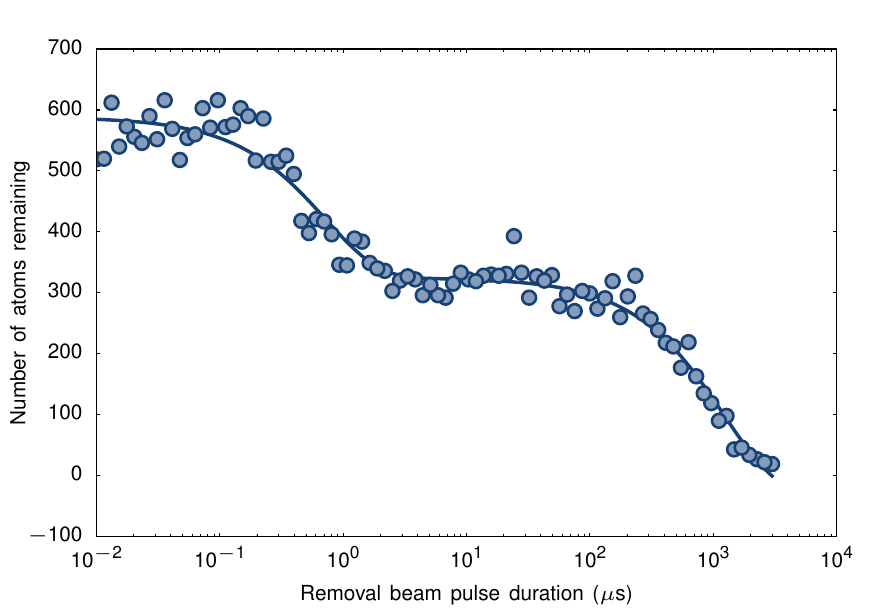}
\caption{Measurement of on- and off-resonant removal of spin states. Here a pulse resonant with removing $\Ket{1}$ is applied for variable time, followed by two $100 \, \mu \mathrm{s}$ pulses resonant with removing $\Ket{4}$ and $\Ket{5}$ (see Spin-removal technique). The remaining atom number as a function of pulse duration is plotted. The data fits well to a sum of two exponentials, yielding an on-resonant characteristic removal time of $0.71(8) \, \mu \mathrm{s}$ and an off-resonant time of $1200(200) \, \mu \mathrm{s}$.}
\label{fig:blowouttime}
\end{figure}

\clearpage

\begin{figure}[h]
\centering
\includegraphics{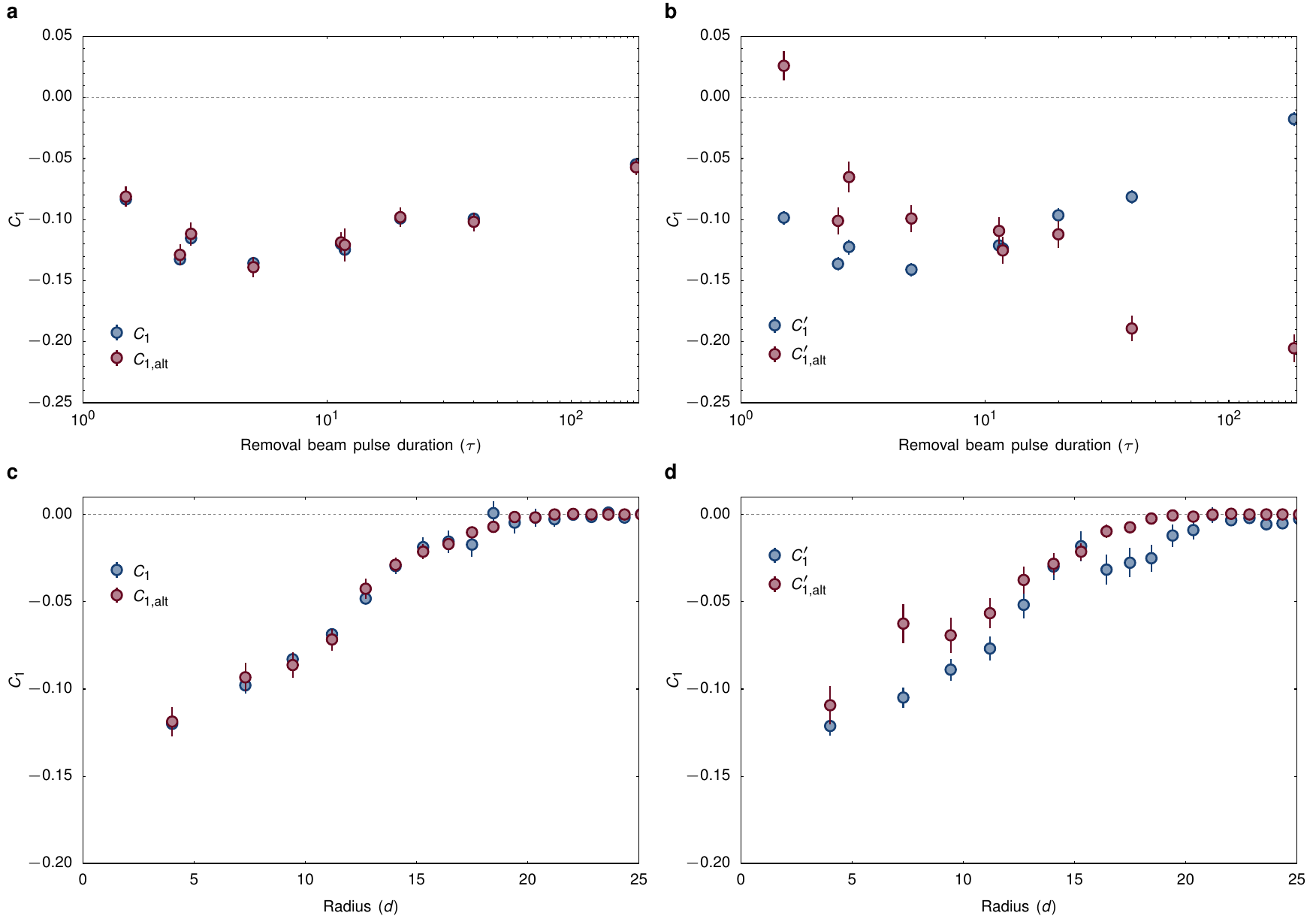}
\caption{Measurement of systematic on correlation due to removal pulse.  (A) The correlator resulting from varying removal pulse duration is plotted for both corrected methods used to measure the spin correlator. The two methods agree for all durations. The reduction in correlations away from the chosen removal time is due to the loss of information caused by incomplete or off-resonant removal. (B) Analogous plot for the uncorrected spin correlators. Without the correction, the correlators agree at our chosen removal time but diverge strongly away from that point. (C, D) Exemplar profiles of the correlator with and without correction, respectively. While the corrected correlators demonstrate agreement across the entire sample, the uncorrected correlators do not. Based on typical fluctuations in spin removal efficiency, for the corrected spin correlator we estimate the systematic error due to spin removal to be $3(1) \times 10^{-4}$.}
\label{fig:blowoutsys}
\end{figure}

\clearpage

\begin{figure}[h]
\centering
\includegraphics{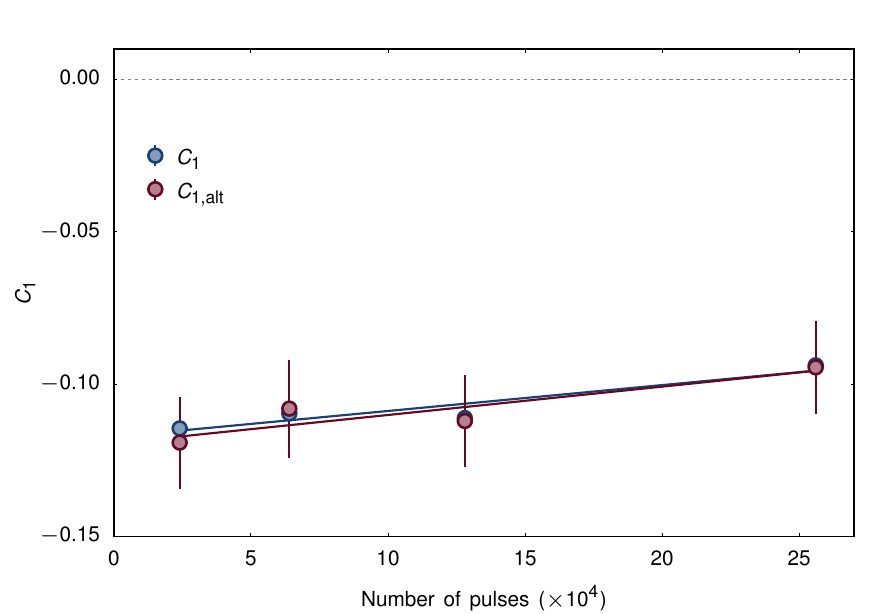}
\caption{Measurement of systematic on correlation due to imaging. For both corrected methods used to measure the correlator, the correlator resulting from a varying number of imaging pulses is plotted. Across the entire range investigated, the two methods agree. Based on fluctuations in imaging efficiency, we estimate the systematic error due to imaging to be $4(2) \times 10^{-5}$ .}
\label{fig:imagingsys}
\end{figure}

\clearpage

\begin{figure}[h]
\centering
\includegraphics{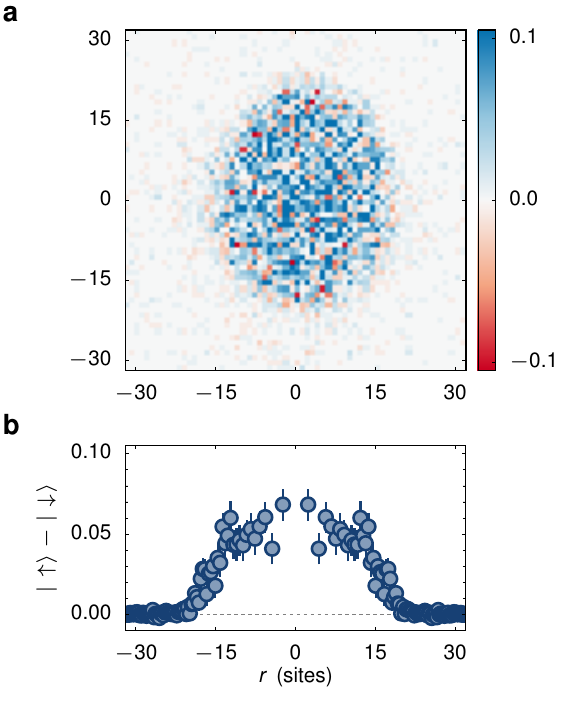}
\caption{Map (A) and profile (B) of difference in spin densities $\Ket{\uparrow} - \Ket{\downarrow}$ over the entire cloud, for the colder data from Figure 2a in the main text. We find a constant polarization $(\Ket{\uparrow} - \Ket{\downarrow})/(\Ket{\uparrow} + \Ket{\downarrow})$ across the entirety of the cloud of about $6.5\%$, and no features in the polarization distribution. This polarization arises from our cloud preparation and loading, and remains the same for all of our data shown in the paper.}
\label{fig:polarization}
\end{figure}

\clearpage

\begin{figure}[h]
\centering
\includegraphics{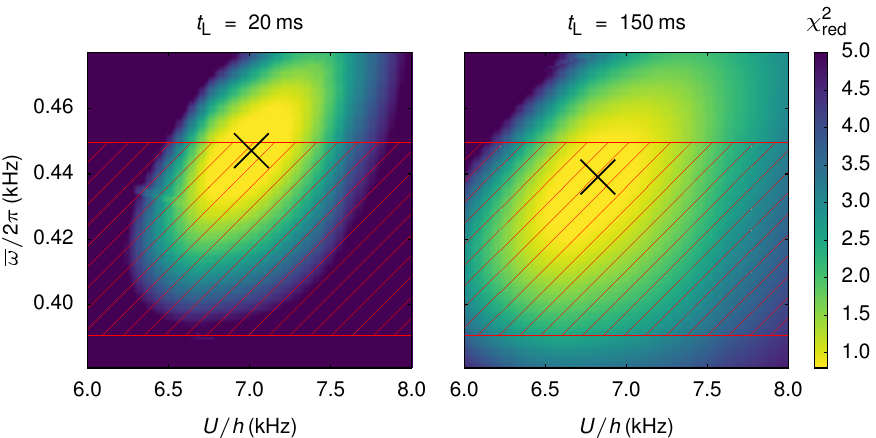}
\caption{Precision determination of the trap frequency and on-site interaction. We show the reduced chi-squared value obtained from simultaneous fits to radial profiles of the density and spin correlator for different pair values of $\nu$ and $U$. We use data shown in Fig. 3 of the main manuscript with two different lattice loading times that are expected to result in samples consistent with thermal equilibrium. Crosses mark the minimum and the red grid denotes the uncertainty region for the trap frequency calibration obtained from breathing oscillations. Using the data for the longest lattice loading time $t_{\mathrm{L}} = 150\, \mathrm{ms}$ we determine $\omega / 2\pi=0.439(5)\,\mathrm{kHz}$ and $U/h=6.82(10)\,\mathrm{kHz}$}
\label{fig:interactionfits}
\end{figure}


\begin{thebibliography}{30}%
\makeatletter
\providecommand \@ifxundefined [1]{%
 \@ifx{#1\undefined}
}%
\providecommand \@ifnum [1]{%
 \ifnum #1\expandafter \@firstoftwo
 \else \expandafter \@secondoftwo
 \fi
}%
\providecommand \@ifx [1]{%
 \ifx #1\expandafter \@firstoftwo
 \else \expandafter \@secondoftwo
 \fi
}%
\providecommand \natexlab [1]{#1}%
\providecommand \enquote  [1]{``#1''}%
\providecommand \bibnamefont  [1]{#1}%
\providecommand \bibfnamefont [1]{#1}%
\providecommand \citenamefont [1]{#1}%
\providecommand \href@noop [0]{\@secondoftwo}%
\providecommand \href [0]{\begingroup \@sanitize@url \@href}%
\providecommand \@href[1]{\@@startlink{#1}\@@href}%
\providecommand \@@href[1]{\endgroup#1\@@endlink}%
\providecommand \@sanitize@url [0]{\catcode `\\12\catcode `\$12\catcode
  `\&12\catcode `\#12\catcode `\^12\catcode `\_12\catcode `\%12\relax}%
\providecommand \@@startlink[1]{}%
\providecommand \@@endlink[0]{}%
\providecommand \url  [0]{\begingroup\@sanitize@url \@url }%
\providecommand \@url [1]{\endgroup\@href {#1}{\urlprefix }}%
\providecommand \urlprefix  [0]{URL }%
\providecommand \Eprint [0]{\href }%
\providecommand \doibase [0]{http://dx.doi.org/}%
\providecommand \selectlanguage [0]{\@gobble}%
\providecommand \bibinfo  [0]{\@secondoftwo}%
\providecommand \bibfield  [0]{\@secondoftwo}%
\providecommand \translation [1]{[#1]}%
\providecommand \BibitemOpen [0]{}%
\providecommand \bibitemStop [0]{}%
\providecommand \bibitemNoStop [0]{.\EOS\space}%
\providecommand \EOS [0]{\spacefactor3000\relax}%
\providecommand \BibitemShut  [1]{\csname bibitem#1\endcsname}%
\let\auto@bib@innerbib\@empty
\expandafter\ifx\csname url\endcsname\relax
  \def\url#1{\texttt{#1}}\fi
\expandafter\ifx\csname urlprefix\endcsname\relax\def\urlprefix{URL }\fi
\providecommand{\bibinfo}[2]{#2}
\providecommand{\eprint}[2][]{\url{#2}}
\expandafter\ifx\csname url\endcsname\relax
  \def\url#1{\texttt{#1}}\fi
\expandafter\ifx\csname urlprefix\endcsname\relax\def\urlprefix{URL }\fi
\providecommand{\bibinfo}[2]{#2}
\providecommand{\eprint}[2][]{\url{#2}}


\bibitem [{\citenamefont {Anderson}(1987)}]{Anderson1987a}%
  \BibitemOpen
  \bibfield  {author} {\bibinfo {author} {\bibfnamefont {P.~W.}\ \bibnamefont
  {Anderson}},\ }\href {\doibase 10.1126/science.235.4793.1196} {\bibfield
  {journal} {\bibinfo  {journal} {Science}\ }\textbf {\bibinfo {volume}
  {235}},\ \bibinfo {pages} {1196} (\bibinfo {year} {1987})}\BibitemShut
  {NoStop}%
\bibitem [{\citenamefont {Lee}\ \emph {et~al.}(2006)\citenamefont {Lee},
  \citenamefont {Nagaosa},\ and\ \citenamefont {Wen}}]{Lee2006}%
  \BibitemOpen
  \bibfield  {author} {\bibinfo {author} {\bibfnamefont {P.}~\bibnamefont
  {Lee}}, \bibinfo {author} {\bibfnamefont {N.}~\bibnamefont {Nagaosa}}, \ and\
  \bibinfo {author} {\bibfnamefont {X.-G.}\ \bibnamefont {Wen}},\ }\href
  {\doibase 10.1103/RevModPhys.78.17} {\bibfield  {journal} {\bibinfo
  {journal} {Rev. Mod. Phys.}\ }\textbf {\bibinfo {volume} {78}},\ \bibinfo
  {pages} {17} (\bibinfo {year} {2006})}\BibitemShut {NoStop}%
\bibitem [{\citenamefont {Esslinger}(2010)}]{Esslinger2010a}%
  \BibitemOpen
  \bibfield  {author} {\bibinfo {author} {\bibfnamefont {T.}~\bibnamefont
  {Esslinger}},\ }\href {\doibase 10.1146/annurev-conmatphys-070909-104059}
  {\bibfield  {journal} {\bibinfo  {journal} {Annu. Rev. Condens. Matter
  Phys.}\ }\textbf {\bibinfo {volume} {1}},\ \bibinfo {pages} {129} (\bibinfo
  {year} {2010})}\BibitemShut {NoStop}%
\bibitem [{\citenamefont {Bloch}\ \emph {et~al.}(2012)\citenamefont {Bloch},
  \citenamefont {Dalibard},\ and\ \citenamefont
  {Nascimb{\`{e}}ne}}]{Bloch2012}%
  \BibitemOpen
  \bibfield  {author} {\bibinfo {author} {\bibfnamefont {I.}~\bibnamefont
  {Bloch}}, \bibinfo {author} {\bibfnamefont {J.}~\bibnamefont {Dalibard}}, \
  and\ \bibinfo {author} {\bibfnamefont {S.}~\bibnamefont {Nascimb{\`{e}}ne}},\
  }\href {\doibase 10.1038/nphys2259} {\bibfield  {journal} {\bibinfo
  {journal} {Nat. Phys.}\ }\textbf {\bibinfo {volume} {8}},\ \bibinfo {pages}
  {267} (\bibinfo {year} {2012})}\BibitemShut {NoStop}%
\bibitem [{\citenamefont {Greif}\ \emph {et~al.}(2013)\citenamefont {Greif},
  \citenamefont {Uehlinger}, \citenamefont {Jotzu}, \citenamefont {Tarruell},\
  and\ \citenamefont {Esslinger}}]{Greif2013a}%
  \BibitemOpen
  \bibfield  {author} {\bibinfo {author} {\bibfnamefont {D.}~\bibnamefont
  {Greif}}, \bibinfo {author} {\bibfnamefont {T.}~\bibnamefont {Uehlinger}},
  \bibinfo {author} {\bibfnamefont {G.}~\bibnamefont {Jotzu}}, \bibinfo
  {author} {\bibfnamefont {L.}~\bibnamefont {Tarruell}}, \ and\ \bibinfo
  {author} {\bibfnamefont {T.}~\bibnamefont {Esslinger}},\ }\href {\doibase
  10.1126/science.1236362} {\bibfield  {journal} {\bibinfo  {journal}
  {Science}\ }\textbf {\bibinfo {volume} {340}},\ \bibinfo {pages} {1307}
  (\bibinfo {year} {2013})}\BibitemShut {NoStop}%
\bibitem [{\citenamefont {Hart}\ \emph {et~al.}(2015)\citenamefont {Hart},
  \citenamefont {Duarte}, \citenamefont {Yang}, \citenamefont {Liu},
  \citenamefont {Paiva}, \citenamefont {Khatami}, \citenamefont {Scalettar},
  \citenamefont {Trivedi}, \citenamefont {Huse},\ and\ \citenamefont
  {Hulet}}]{Hart2014}%
  \BibitemOpen
  \bibfield  {author} {\bibinfo {author} {\bibfnamefont {R.~A.}\ \bibnamefont
  {Hart}}, \bibinfo {author} {\bibfnamefont {P.~M.}\ \bibnamefont {Duarte}},
  \bibinfo {author} {\bibfnamefont {T.-L.}\ \bibnamefont {Yang}}, \bibinfo
  {author} {\bibfnamefont {X.}~\bibnamefont {Liu}}, \bibinfo {author}
  {\bibfnamefont {T.}~\bibnamefont {Paiva}}, \bibinfo {author} {\bibfnamefont
  {E.}~\bibnamefont {Khatami}}, \bibinfo {author} {\bibfnamefont {R.~T.}\
  \bibnamefont {Scalettar}}, \bibinfo {author} {\bibfnamefont {N.}~\bibnamefont
  {Trivedi}}, \bibinfo {author} {\bibfnamefont {D.~A.}\ \bibnamefont {Huse}}, \
  and\ \bibinfo {author} {\bibfnamefont {R.~G.}\ \bibnamefont {Hulet}},\ }\href
  {\doibase 10.1038/nature14223} {\bibfield  {journal} {\bibinfo  {journal}
  {Nature}\ }\textbf {\bibinfo {volume} {519}},\ \bibinfo {pages} {211}
  (\bibinfo {year} {2015})}\BibitemShut {NoStop}%
\bibitem [{\citenamefont {Haller}\ \emph {et~al.}(2015)\citenamefont {Haller},
  \citenamefont {Hudson}, \citenamefont {Kelly}, \citenamefont {Cotta},
  \citenamefont {Peaudecerf}, \citenamefont {Bruce},\ and\ \citenamefont
  {Kuhr}}]{Haller2015a}%
  \BibitemOpen
  \bibfield  {author} {\bibinfo {author} {\bibfnamefont {E.}~\bibnamefont
  {Haller}}, \bibinfo {author} {\bibfnamefont {J.}~\bibnamefont {Hudson}},
  \bibinfo {author} {\bibfnamefont {A.}~\bibnamefont {Kelly}}, \bibinfo
  {author} {\bibfnamefont {D.~A.}\ \bibnamefont {Cotta}}, \bibinfo {author}
  {\bibfnamefont {B.}~\bibnamefont {Peaudecerf}}, \bibinfo {author}
  {\bibfnamefont {G.~D.}\ \bibnamefont {Bruce}}, \ and\ \bibinfo {author}
  {\bibfnamefont {S.}~\bibnamefont {Kuhr}},\ }\href {\doibase 10.1038/nphys3403} 
  {\bibfield  {journal} {\bibinfo  {journal} {Nat. Phys.}\
  }\textbf {\bibinfo {volume} {11}},\ \bibinfo {pages} {738} (\bibinfo {year}
  {2015})}\BibitemShut {NoStop}%
\bibitem [{\citenamefont {Cheuk}\ \emph {et~al.}(2015)\citenamefont {Cheuk},
  \citenamefont {Nichols}, \citenamefont {Okan}, \citenamefont {Gersdorf},
  \citenamefont {Ramasesh}, \citenamefont {Bakr}, \citenamefont {Lompe},\ and\
  \citenamefont {Zwierlein}}]{Cheuk}%
  \BibitemOpen
  \bibfield  {author} {\bibinfo {author} {\bibfnamefont {L.~W.}\ \bibnamefont
  {Cheuk}}, \bibinfo {author} {\bibfnamefont {M.~A.}\ \bibnamefont {Nichols}},
  \bibinfo {author} {\bibfnamefont {M.}~\bibnamefont {Okan}}, \bibinfo {author}
  {\bibfnamefont {T.}~\bibnamefont {Gersdorf}}, \bibinfo {author}
  {\bibfnamefont {V.~V.}\ \bibnamefont {Ramasesh}}, \bibinfo {author}
  {\bibfnamefont {W.~S.}\ \bibnamefont {Bakr}}, \bibinfo {author}
  {\bibfnamefont {T.}~\bibnamefont {Lompe}}, \ and\ \bibinfo {author}
  {\bibfnamefont {M.~W.}\ \bibnamefont {Zwierlein}},\ }\href {\doibase 10.1103/PhysRevLett.114.193001} 
  {\bibfield  {journal} {\bibinfo  {journal}
  {Phys. Rev. Lett.}\ }\textbf {\bibinfo {volume} {114}},\ \bibinfo {pages}
  {193001} (\bibinfo {year} {2015})}\BibitemShut {NoStop}%
\bibitem [{\citenamefont {Parsons}\ \emph {et~al.}(2015)\citenamefont
  {Parsons}, \citenamefont {Huber}, \citenamefont {Mazurenko}, \citenamefont
  {Chiu}, \citenamefont {Setiawan}, \citenamefont {Wooley-Brown}, \citenamefont
  {Blatt},\ and\ \citenamefont {Greiner}}]{Parsons2015}%
  \BibitemOpen
  \bibfield  {author} {\bibinfo {author} {\bibfnamefont {M.~F.}\ \bibnamefont
  {Parsons}}, \bibinfo {author} {\bibfnamefont {F.}~\bibnamefont {Huber}},
  \bibinfo {author} {\bibfnamefont {A.}~\bibnamefont {Mazurenko}}, \bibinfo
  {author} {\bibfnamefont {C.~S.}\ \bibnamefont {Chiu}}, \bibinfo {author}
  {\bibfnamefont {W.}~\bibnamefont {Setiawan}}, \bibinfo {author}
  {\bibfnamefont {K.}~\bibnamefont {Wooley-Brown}}, \bibinfo {author}
  {\bibfnamefont {S.}~\bibnamefont {Blatt}}, \ and\ \bibinfo {author}
  {\bibfnamefont {M.}~\bibnamefont {Greiner}},\ }\href {\doibase 10.1103/PhysRevLett.114.213002} 
  {\bibfield  {journal} {\bibinfo  {journal}
  {Phys. Rev. Lett.}\ }\textbf {\bibinfo {volume} {114}},\ \bibinfo {pages}
  {213002} (\bibinfo {year} {2015})}\BibitemShut {NoStop}%
\bibitem [{\citenamefont {Edge}\ \emph {et~al.}(2015)\citenamefont {Edge},
  \citenamefont {Anderson}, \citenamefont {Jervis}, \citenamefont {McKay},
  \citenamefont {Day}, \citenamefont {Trotzky},\ and\ \citenamefont
  {Thywissen}}]{Edge2015}%
  \BibitemOpen
  \bibfield  {author} {\bibinfo {author} {\bibfnamefont {G.~J.~A.}\
  \bibnamefont {Edge}}, \bibinfo {author} {\bibfnamefont {R.}~\bibnamefont
  {Anderson}}, \bibinfo {author} {\bibfnamefont {D.}~\bibnamefont {Jervis}},
  \bibinfo {author} {\bibfnamefont {D.~C.}\ \bibnamefont {McKay}}, \bibinfo
  {author} {\bibfnamefont {R.}~\bibnamefont {Day}}, \bibinfo {author}
  {\bibfnamefont {S.}~\bibnamefont {Trotzky}}, \ and\ \bibinfo {author}
  {\bibfnamefont {J.~H.}\ \bibnamefont {Thywissen}},\ }\href {\doibase 10.1103/PhysRevA.92.063406} 
  {\bibfield  {journal} {\bibinfo  {journal} {Phys.
  Rev. A}\ }\textbf {\bibinfo {volume} {92}},\ \bibinfo {pages} {063406}
  (\bibinfo {year} {2015})}\BibitemShut {NoStop}%
\bibitem [{\citenamefont {Omran}\ \emph {et~al.}(2015)\citenamefont {Omran},
  \citenamefont {Boll}, \citenamefont {Hilker}, \citenamefont {Kleinlein},
  \citenamefont {Salomon}, \citenamefont {Bloch},\ and\ \citenamefont
  {Gross}}]{Omran2015}%
  \BibitemOpen
  \bibfield  {author} {\bibinfo {author} {\bibfnamefont {A.}~\bibnamefont
  {Omran}}, \bibinfo {author} {\bibfnamefont {M.}~\bibnamefont {Boll}},
  \bibinfo {author} {\bibfnamefont {T.~A.}\ \bibnamefont {Hilker}}, \bibinfo
  {author} {\bibfnamefont {K.}~\bibnamefont {Kleinlein}}, \bibinfo {author}
  {\bibfnamefont {G.}~\bibnamefont {Salomon}}, \bibinfo {author} {\bibfnamefont
  {I.}~\bibnamefont {Bloch}}, \ and\ \bibinfo {author} {\bibfnamefont
  {C.}~\bibnamefont {Gross}},\ }\href {\doibase 10.1103/PhysRevLett.115.263001}
  {\bibfield  {journal} {\bibinfo  {journal} {Phys. Rev. Lett.}\ }\textbf
  {\bibinfo {volume} {115}},\ \bibinfo {pages} {263001} (\bibinfo {year}
  {2015})}\BibitemShut {NoStop}%
\bibitem [{\citenamefont {Greif}\ \emph {et~al.}(2016)\citenamefont {Greif},
  \citenamefont {Parsons}, \citenamefont {Mazurenko}, \citenamefont {Chiu},
  \citenamefont {Blatt}, \citenamefont {Huber}, \citenamefont {Ji},\ and\
  \citenamefont {Greiner}}]{Greif2015}%
  \BibitemOpen
  \bibfield  {author} {\bibinfo {author} {\bibfnamefont {D.}~\bibnamefont
  {Greif}}, \bibinfo {author} {\bibfnamefont {M.~F.}\ \bibnamefont {Parsons}},
  \bibinfo {author} {\bibfnamefont {A.}~\bibnamefont {Mazurenko}}, \bibinfo
  {author} {\bibfnamefont {C.~S.}\ \bibnamefont {Chiu}}, \bibinfo {author}
  {\bibfnamefont {S.}~\bibnamefont {Blatt}}, \bibinfo {author} {\bibfnamefont
  {F.}~\bibnamefont {Huber}}, \bibinfo {author} {\bibfnamefont
  {G.}~\bibnamefont {Ji}}, \ and\ \bibinfo {author} {\bibfnamefont
  {M.}~\bibnamefont {Greiner}},\ }\href {\doibase 10.1126/science.aad9041}
  {\bibfield  {journal} {\bibinfo  {journal} {Science}\ }\textbf {\bibinfo
  {volume} {351}},\ \bibinfo {pages} {953} (\bibinfo {year}
  {2016})}\BibitemShut {NoStop}%
\bibitem [{\citenamefont {Cheuk}\ \emph {et~al.}(2016)\citenamefont {Cheuk},
  \citenamefont {Nichols}, \citenamefont {Lawrence}, \citenamefont {Okan},
  \citenamefont {Zhang},\ and\ \citenamefont {Zwierlein}}]{Cheuk2016}%
  \BibitemOpen
  \bibfield  {author} {\bibinfo {author} {\bibfnamefont {L.~W.}\ \bibnamefont
  {Cheuk}}, \bibinfo {author} {\bibfnamefont {M.~A.}\ \bibnamefont {Nichols}},
  \bibinfo {author} {\bibfnamefont {K.~R.}\ \bibnamefont {Lawrence}}, \bibinfo
  {author} {\bibfnamefont {M.}~\bibnamefont {Okan}}, \bibinfo {author}
  {\bibfnamefont {H.}~\bibnamefont {Zhang}}, \ and\ \bibinfo {author}
  {\bibfnamefont {M.~W.}\ \bibnamefont {Zwierlein}},\ }\href {\doibase 10.1103/PhysRevLett.116.235301} 
  {\bibfield  {journal} {\bibinfo  {journal}
  {Phys. Rev. Lett.}\ }\textbf {\bibinfo {volume} {116}},\ \bibinfo {pages}
  {235301} (\bibinfo {year} {2016})}\BibitemShut {NoStop}%
\bibitem [{\citenamefont {Bakr}\ \emph {et~al.}(2010)\citenamefont {Bakr},
  \citenamefont {Peng}, \citenamefont {Tai}, \citenamefont {Ma}, \citenamefont
  {Simon}, \citenamefont {Gillen}, \citenamefont {Foelling}, \citenamefont
  {Pollet},\ and\ \citenamefont {Greiner}}]{Bakr2010}%
  \BibitemOpen
  \bibfield  {author} {\bibinfo {author} {\bibfnamefont {W.~S.}\ \bibnamefont
  {Bakr}}, \bibinfo {author} {\bibfnamefont {A.}~\bibnamefont {Peng}}, \bibinfo
  {author} {\bibfnamefont {M.~E.}\ \bibnamefont {Tai}}, \bibinfo {author}
  {\bibfnamefont {R.}~\bibnamefont {Ma}}, \bibinfo {author} {\bibfnamefont
  {J.}~\bibnamefont {Simon}}, \bibinfo {author} {\bibfnamefont {J.~I.}\
  \bibnamefont {Gillen}}, \bibinfo {author} {\bibfnamefont {S.}~\bibnamefont
  {Foelling}}, \bibinfo {author} {\bibfnamefont {L.}~\bibnamefont {Pollet}}, \
  and\ \bibinfo {author} {\bibfnamefont {M.}~\bibnamefont {Greiner}},\ }\href
  {http://science.sciencemag.org/content/329/5991/547} {\bibfield  {journal}
  {\bibinfo  {journal} {Science}\ }\textbf {\bibinfo {volume} {329}},\ \bibinfo
  {pages} {547} (\bibinfo {year} {2010})}\BibitemShut {NoStop}%
\bibitem [{\citenamefont {Sherson}\ \emph {et~al.}(2010)\citenamefont
  {Sherson}, \citenamefont {Weitenberg}, \citenamefont {Endres}, \citenamefont
  {Cheneau}, \citenamefont {Bloch},\ and\ \citenamefont {Kuhr}}]{Sherson2010}%
  \BibitemOpen
  \bibfield  {author} {\bibinfo {author} {\bibfnamefont {J.~F.}\ \bibnamefont
  {Sherson}}, \bibinfo {author} {\bibfnamefont {C.}~\bibnamefont {Weitenberg}},
  \bibinfo {author} {\bibfnamefont {M.}~\bibnamefont {Endres}}, \bibinfo
  {author} {\bibfnamefont {M.}~\bibnamefont {Cheneau}}, \bibinfo {author}
  {\bibfnamefont {I.}~\bibnamefont {Bloch}}, \ and\ \bibinfo {author}
  {\bibfnamefont {S.}~\bibnamefont {Kuhr}},\ }\href {\doibase 10.1038/nature09378} 
  {\bibfield  {journal} {\bibinfo  {journal} {Nature}\
  }\textbf {\bibinfo {volume} {467}},\ \bibinfo {pages} {68} (\bibinfo {year}
  {2010})}\BibitemShut {NoStop}%
\bibitem [{\citenamefont {Cocchi}\ \emph {et~al.}(2016)\citenamefont {Cocchi},
  \citenamefont {Miller}, \citenamefont {Drewes}, \citenamefont {Pertot},
  \citenamefont {Brennecke},\ and\ \citenamefont {Michael}}]{Cocchi2016}%
  \BibitemOpen
  \bibfield  {author} {\bibinfo {author} {\bibfnamefont {E.}~\bibnamefont
  {Cocchi}}, \bibinfo {author} {\bibfnamefont {L.~A.}\ \bibnamefont {Miller}},
  \bibinfo {author} {\bibfnamefont {J.~H.}\ \bibnamefont {Drewes}}, \bibinfo
  {author} {\bibfnamefont {D.}~\bibnamefont {Pertot}}, \bibinfo {author}
  {\bibfnamefont {F.}~\bibnamefont {Brennecke}}, \ and\ \bibinfo {author}
  {\bibfnamefont {K.}~\bibnamefont {Michael}},\ }\href {\doibase 10.1103/PhysRevLett.116.175301} 
  {\bibfield  {journal} {\bibinfo  {journal}
  {Phys. Rev. Lett.}\ }\textbf {\bibinfo {volume} {116}},\ \bibinfo {pages}
  {175301} (\bibinfo {year} {2016})}\BibitemShut {NoStop}%
\bibitem [{\citenamefont {Preiss}\ \emph {et~al.}(2015)\citenamefont {Preiss},
  \citenamefont {Tai}, \citenamefont {Lukin}, \citenamefont {Rispoli},
  \citenamefont {Zupancic}, \citenamefont {Lahini}, \citenamefont {Islam},\
  and\ \citenamefont {Greiner}}]{Preiss}%
  \BibitemOpen
  \bibfield  {author} {\bibinfo {author} {\bibfnamefont {P.~M.}\ \bibnamefont
  {Preiss}}, \bibinfo {author} {\bibfnamefont {M.~E.}\ \bibnamefont {Tai}},
  \bibinfo {author} {\bibfnamefont {A.}~\bibnamefont {Lukin}}, \bibinfo
  {author} {\bibfnamefont {M.}~\bibnamefont {Rispoli}}, \bibinfo {author}
  {\bibfnamefont {P.}~\bibnamefont {Zupancic}}, \bibinfo {author}
  {\bibfnamefont {Y.}~\bibnamefont {Lahini}}, \bibinfo {author} {\bibfnamefont
  {R.}~\bibnamefont {Islam}}, \ and\ \bibinfo {author} {\bibfnamefont
  {M.}~\bibnamefont {Greiner}},\ }\href {\doibase 10.1126/science.1260364}
  {\bibfield  {journal} {\bibinfo  {journal} {Science}\ }\textbf {\bibinfo
  {volume} {347}},\ \bibinfo {pages} {1229} (\bibinfo {year}
  {2015})}\BibitemShut {NoStop}%
\bibitem [{\citenamefont {Islam}\ \emph {et~al.}(2015)\citenamefont {Islam},
  \citenamefont {Ma}, \citenamefont {Preiss}, \citenamefont {{Eric Tai}},
  \citenamefont {Lukin}, \citenamefont {Rispoli},\ and\ \citenamefont
  {Greiner}}]{Islam2015}%
  \BibitemOpen
  \bibfield  {author} {\bibinfo {author} {\bibfnamefont {R.}~\bibnamefont
  {Islam}}, \bibinfo {author} {\bibfnamefont {R.}~\bibnamefont {Ma}}, \bibinfo
  {author} {\bibfnamefont {P.~M.}\ \bibnamefont {Preiss}}, \bibinfo {author}
  {\bibfnamefont {M.}~\bibnamefont {{Eric Tai}}}, \bibinfo {author}
  {\bibfnamefont {A.}~\bibnamefont {Lukin}}, \bibinfo {author} {\bibfnamefont
  {M.}~\bibnamefont {Rispoli}}, \ and\ \bibinfo {author} {\bibfnamefont
  {M.}~\bibnamefont {Greiner}},\ }\href {\doibase 10.1038/nature15750}
  {\bibfield  {journal} {\bibinfo  {journal} {Nature}\ }\textbf {\bibinfo
  {volume} {528}},\ \bibinfo {pages} {77} (\bibinfo {year} {2015})}\BibitemShut
  {NoStop}%
\bibitem [{\citenamefont {Fukuhara}\ \emph {et~al.}(2013)\citenamefont
  {Fukuhara}, \citenamefont {Schau{\ss}}, \citenamefont {Endres}, \citenamefont
  {Hild}, \citenamefont {Cheneau}, \citenamefont {Bloch},\ and\ \citenamefont
  {Gross}}]{Fukuhara2013}%
  \BibitemOpen
  \bibfield  {author} {\bibinfo {author} {\bibfnamefont {T.}~\bibnamefont
  {Fukuhara}}, \bibinfo {author} {\bibfnamefont {P.}~\bibnamefont
  {Schau{\ss}}}, \bibinfo {author} {\bibfnamefont {M.}~\bibnamefont {Endres}},
  \bibinfo {author} {\bibfnamefont {S.}~\bibnamefont {Hild}}, \bibinfo {author}
  {\bibfnamefont {M.}~\bibnamefont {Cheneau}}, \bibinfo {author} {\bibfnamefont
  {I.}~\bibnamefont {Bloch}}, \ and\ \bibinfo {author} {\bibfnamefont
  {C.}~\bibnamefont {Gross}},\ }\href {\doibase 10.1038/nature12541} {\bibfield
   {journal} {\bibinfo  {journal} {Nature}\ }\textbf {\bibinfo {volume}
  {502}},\ \bibinfo {pages} {76} (\bibinfo {year} {2013})}\BibitemShut
  {NoStop}%
\bibitem [{\citenamefont {Hild}\ \emph {et~al.}(2014)\citenamefont {Hild},
  \citenamefont {Fukuhara}, \citenamefont {Schau{\ss}}, \citenamefont {Zeiher},
  \citenamefont {Knap}, \citenamefont {Demler}, \citenamefont {Bloch},\ and\
  \citenamefont {Gross}}]{Hild2014}%
  \BibitemOpen
  \bibfield  {author} {\bibinfo {author} {\bibfnamefont {S.}~\bibnamefont
  {Hild}}, \bibinfo {author} {\bibfnamefont {T.}~\bibnamefont {Fukuhara}},
  \bibinfo {author} {\bibfnamefont {P.}~\bibnamefont {Schau{\ss}}}, \bibinfo
  {author} {\bibfnamefont {J.}~\bibnamefont {Zeiher}}, \bibinfo {author}
  {\bibfnamefont {M.}~\bibnamefont {Knap}}, \bibinfo {author} {\bibfnamefont
  {E.}~\bibnamefont {Demler}}, \bibinfo {author} {\bibfnamefont
  {I.}~\bibnamefont {Bloch}}, \ and\ \bibinfo {author} {\bibfnamefont
  {C.}~\bibnamefont {Gross}},\ }\href {\doibase 10.1103/PhysRevLett.113.147205}
  {\bibfield  {journal} {\bibinfo  {journal} {Phys. Rev. Lett.}\ }\textbf
  {\bibinfo {volume} {113}},\ \bibinfo {pages} {147205} (\bibinfo {year}
  {2014})}\BibitemShut {NoStop}%
\bibitem [{\citenamefont {Mermin}\ and\ \citenamefont
  {Wagner}(1966)}]{Mermin1966}%
  \BibitemOpen
  \bibfield  {author} {\bibinfo {author} {\bibfnamefont {N.~D.}\ \bibnamefont
  {Mermin}}\ and\ \bibinfo {author} {\bibfnamefont {H.}~\bibnamefont
  {Wagner}},\ }\href {\doibase 10.1103/PhysRevLett.17.1133} {\bibfield
  {journal} {\bibinfo  {journal} {Phys. Rev. Lett.}\ }\textbf {\bibinfo
  {volume} {17}},\ \bibinfo {pages} {1133} (\bibinfo {year}
  {1966})}\BibitemShut {NoStop}%
\bibitem [{\citenamefont {Z{\"{u}}rn}\ \emph {et~al.}(2013)\citenamefont
  {Z{\"{u}}rn}, \citenamefont {Lompe}, \citenamefont {Wenz}, \citenamefont
  {Jochim}, \citenamefont {Julienne},\ and\ \citenamefont {Hutson}}]{Zurn2013}%
  \BibitemOpen
  \bibfield  {author} {\bibinfo {author} {\bibfnamefont {G.}~\bibnamefont
  {Z{\"{u}}rn}}, \bibinfo {author} {\bibfnamefont {T.}~\bibnamefont {Lompe}},
  \bibinfo {author} {\bibfnamefont {A.~N.}\ \bibnamefont {Wenz}}, \bibinfo
  {author} {\bibfnamefont {S.}~\bibnamefont {Jochim}}, \bibinfo {author}
  {\bibfnamefont {P.~S.}\ \bibnamefont {Julienne}}, \ and\ \bibinfo {author}
  {\bibfnamefont {J.~M.}\ \bibnamefont {Hutson}},\ }\href {\doibase 10.1103/PhysRevLett.110.135301} 
  {\bibfield  {journal} {\bibinfo  {journal}
  {Phys. Rev. Lett.}\ }\textbf {\bibinfo {volume} {110}},\ \bibinfo {pages}
  {135301} (\bibinfo {year} {2013})}\BibitemShut {NoStop}%
\bibitem [{\citenamefont {{S}ee~supplementary material}()}]{supplementary}%
  \BibitemOpen
  \bibfield  {author} {\bibinfo {author} {\bibnamefont {{S}ee~supplementary
  material}},\ }\href@noop {} {}\BibitemShut {NoStop}%
\bibitem [{\citenamefont {Khatami}\ and\ \citenamefont
  {Rigol}(2011)}]{Khatami2011}%
  \BibitemOpen
  \bibfield  {author} {\bibinfo {author} {\bibfnamefont {E.}~\bibnamefont
  {Khatami}}\ and\ \bibinfo {author} {\bibfnamefont {M.}~\bibnamefont
  {Rigol}},\ }\href {\doibase 10.1103/PhysRevA.84.053611} {\bibfield  {journal}
  {\bibinfo  {journal} {Phys. Rev. A}\ }\textbf {\bibinfo {volume} {84}},\
  \bibinfo {pages} {053611} (\bibinfo {year} {2011})}\BibitemShut {NoStop}%
\bibitem [{\citenamefont {Chiesa}\ \emph {et~al.}(2011)\citenamefont {Chiesa},
  \citenamefont {Varney}, \citenamefont {Rigol},\ and\ \citenamefont
  {Scalettar}}]{Chiesa2011}%
  \BibitemOpen
  \bibfield  {author} {\bibinfo {author} {\bibfnamefont {S.}~\bibnamefont
  {Chiesa}}, \bibinfo {author} {\bibfnamefont {C.~N.}\ \bibnamefont {Varney}},
  \bibinfo {author} {\bibfnamefont {M.}~\bibnamefont {Rigol}}, \ and\ \bibinfo
  {author} {\bibfnamefont {R.~T.}\ \bibnamefont {Scalettar}},\ }\href {\doibase 10.1103/PhysRevLett.106.035301} 
  {\bibfield  {journal} {\bibinfo  {journal}
  {Phys. Rev. Lett.}\ }\textbf {\bibinfo {volume} {106}},\ \bibinfo {pages}
  {035301} (\bibinfo {year} {2011})}\BibitemShut {NoStop}%
\bibitem [{\citenamefont {Paiva}\ \emph {et~al.}(2010)\citenamefont {Paiva},
  \citenamefont {Scalettar}, \citenamefont {Randeria},\ and\ \citenamefont
  {Trivedi}}]{Paiva2010}%
  \BibitemOpen
  \bibfield  {author} {\bibinfo {author} {\bibfnamefont {T.}~\bibnamefont
  {Paiva}}, \bibinfo {author} {\bibfnamefont {R.}~\bibnamefont {Scalettar}},
  \bibinfo {author} {\bibfnamefont {M.}~\bibnamefont {Randeria}}, \ and\
  \bibinfo {author} {\bibfnamefont {N.}~\bibnamefont {Trivedi}},\ }\href
  {\doibase 10.1103/PhysRevLett.104.066406} {\bibfield  {journal} {\bibinfo
  {journal} {Phys. Rev. Lett.}\ }\textbf {\bibinfo {volume} {104}},\ \bibinfo
  {pages} {066406} (\bibinfo {year} {2010})}\BibitemShut {NoStop}%
\bibitem [{\citenamefont {B{\"{u}}chler}(2010)}]{Buchler2010}%
  \BibitemOpen
  \bibfield  {author} {\bibinfo {author} {\bibfnamefont {H.~P.}\ \bibnamefont
  {B{\"{u}}chler}},\ }\href {\doibase 10.1103/PhysRevLett.104.090402}
  {\bibfield  {journal} {\bibinfo  {journal} {Phys. Rev. Lett.}\ }\textbf
  {\bibinfo {volume} {104}},\ \bibinfo {pages} {090402} (\bibinfo {year}
  {2010})}\BibitemShut {NoStop}%
\bibitem [{\citenamefont {Bernier}\ \emph {et~al.}(2009)\citenamefont
  {Bernier}, \citenamefont {Kollath}, \citenamefont {Georges}, \citenamefont
  {{De Leo}}, \citenamefont {Gerbier}, \citenamefont {Salomon},\ and\
  \citenamefont {K{\"{o}}hl}}]{Bernier2009}%
  \BibitemOpen
  \bibfield  {author} {\bibinfo {author} {\bibfnamefont {J.-S.}\ \bibnamefont
  {Bernier}}, \bibinfo {author} {\bibfnamefont {C.}~\bibnamefont {Kollath}},
  \bibinfo {author} {\bibfnamefont {A.}~\bibnamefont {Georges}}, \bibinfo
  {author} {\bibfnamefont {L.}~\bibnamefont {{De Leo}}}, \bibinfo {author}
  {\bibfnamefont {F.}~\bibnamefont {Gerbier}}, \bibinfo {author} {\bibfnamefont
  {C.}~\bibnamefont {Salomon}}, \ and\ \bibinfo {author} {\bibfnamefont
  {M.}~\bibnamefont {K{\"{o}}hl}},\ }\href {\doibase 10.1103/PhysRevA.79.061601} 
  {\bibfield  {journal} {\bibinfo  {journal} {Phys.
  Rev. A}\ }\textbf {\bibinfo {volume} {79}},\ \bibinfo {pages} {061601(R)}
  (\bibinfo {year} {2009})}\BibitemShut {NoStop}%
\bibitem [{\citenamefont {Ho}\ and\ \citenamefont {Zhou}(2009)}]{Ho2009b}%
  \BibitemOpen
  \bibfield  {author} {\bibinfo {author} {\bibfnamefont {T.-L.}\ \bibnamefont
  {Ho}}\ and\ \bibinfo {author} {\bibfnamefont {Q.}~\bibnamefont {Zhou}},\
  }\href {\doibase 10.1073/pnas.0809862105} {\bibfield  {journal} {\bibinfo
  {journal} {Proc. Natl. Acad. Sci.}\ }\textbf {\bibinfo {volume} {106}},\
  \bibinfo {pages} {6916} (\bibinfo {year} {2009})}\BibitemShut {NoStop}%
\bibitem [{\citenamefont {Zupancic}\ \emph {et~al.}(2016)\citenamefont
  {Zupancic}, \citenamefont {Preiss}, \citenamefont {Ma}, \citenamefont
  {Lukin}, \citenamefont {Tai}, \citenamefont {Rispoli}, \citenamefont
  {Islam},\ and\ \citenamefont {Greiner}}]{Zupancic2016}%
  \BibitemOpen
  \bibfield  {author} {\bibinfo {author} {\bibfnamefont {P.}~\bibnamefont
  {Zupancic}}, \bibinfo {author} {\bibfnamefont {P.~M.}\ \bibnamefont
  {Preiss}}, \bibinfo {author} {\bibfnamefont {R.}~\bibnamefont {Ma}}, \bibinfo
  {author} {\bibfnamefont {A.}~\bibnamefont {Lukin}}, \bibinfo {author}
  {\bibfnamefont {M.~E.}\ \bibnamefont {Tai}}, \bibinfo {author} {\bibfnamefont
  {M.}~\bibnamefont {Rispoli}}, \bibinfo {author} {\bibfnamefont
  {R.}~\bibnamefont {Islam}}, \ and\ \bibinfo {author} {\bibfnamefont
  {M.}~\bibnamefont {Greiner}},\ }\href {\doibase 10.1364/OE.24.013881}
  {\bibfield  {journal} {\bibinfo  {journal} {Opt. Express}\ }\textbf {\bibinfo
  {volume} {24}},\ \bibinfo {pages} {13881} (\bibinfo {year}
  {2016})}\BibitemShut {NoStop}%
\bibitem [{\citenamefont {Macridin}\ \emph {et~al.}(2006)\citenamefont
  {Macridin}, \citenamefont {Jarrell}, \citenamefont {Maier}, \citenamefont
  {Kent},\ and\ \citenamefont {D'Azevedo}}]{Macridin2006}%
  \BibitemOpen
  \bibfield  {author} {\bibinfo {author} {\bibfnamefont {A.}~\bibnamefont
  {Macridin}}, \bibinfo {author} {\bibfnamefont {M.}~\bibnamefont {Jarrell}},
  \bibinfo {author} {\bibfnamefont {T.}~\bibnamefont {Maier}}, \bibinfo
  {author} {\bibfnamefont {P.~R.~C.}\ \bibnamefont {Kent}}, \ and\ \bibinfo
  {author} {\bibfnamefont {E.}~\bibnamefont {D'Azevedo}},\ }\href {\doibase 10.1103/PhysRevLett.97.036401} 
  {\bibfield  {journal} {\bibinfo  {journal}
  {Phys. Rev. Lett.}\ }\textbf {\bibinfo {volume} {97}},\ \bibinfo {pages}
  {036401} (\bibinfo {year} {2006})}\BibitemShut {NoStop}%
\bibitem [{\citenamefont {Gull}\ \emph {et~al.}(2013)\citenamefont {Gull},
  \citenamefont {Parcollet},\ and\ \citenamefont {Millis}}]{Gull2013}%
  \BibitemOpen
  \bibfield  {author} {\bibinfo {author} {\bibfnamefont {E.}~\bibnamefont
  {Gull}}, \bibinfo {author} {\bibfnamefont {O.}~\bibnamefont {Parcollet}}, \
  and\ \bibinfo {author} {\bibfnamefont {A.~J.}\ \bibnamefont {Millis}},\
  }\href {\doibase 10.1103/PhysRevLett.110.216405} {\bibfield  {journal}
  {\bibinfo  {journal} {Phys. Rev. Lett.}\ }\textbf {\bibinfo {volume} {110}},\
  \bibinfo {pages} {1} (\bibinfo {year} {2013})}\BibitemShut {NoStop}%
\bibitem [{\citenamefont {Eisert}\ \emph {et~al.}(2015)\citenamefont {Eisert},
  \citenamefont {Friesdorf},\ and\ \citenamefont {Gogolin}}]{Eisert2014}%
  \BibitemOpen
  \bibfield  {author} {\bibinfo {author} {\bibfnamefont {J.}~\bibnamefont
  {Eisert}}, \bibinfo {author} {\bibfnamefont {M.}~\bibnamefont {Friesdorf}}, \
  and\ \bibinfo {author} {\bibfnamefont {C.}~\bibnamefont {Gogolin}},\ }\href
  {\doibase 10.1038/nphys3215} {\bibfield  {journal} {\bibinfo  {journal} {Nat.
  Phys.}\ }\textbf {\bibinfo {volume} {11}},\ \bibinfo {pages} {124} (\bibinfo
  {year} {2015})}\BibitemShut {NoStop}%
\bibitem [{\citenamefont {van~der Walt}\ \emph {et~al.}(2014)\citenamefont
  {van~der Walt}, \citenamefont {{S}ch\"onberger}, \citenamefont
  {{Nunez-Iglesias}}, \citenamefont {{B}oulogne}, \citenamefont {{W}arner},
  \citenamefont {{Y}ager}, \citenamefont {{G}ouillart}, \citenamefont {{Y}u},\
  and\ \citenamefont {the scikit-image contributors}}]{scikit-image}%
  \BibitemOpen
  \bibfield  {author} {\bibinfo {author} {\bibfnamefont {S.}~\bibnamefont
  {van~der Walt}}, \bibinfo {author} {\bibfnamefont {J.~L.}\ \bibnamefont
  {{S}ch\"onberger}}, \bibinfo {author} {\bibfnamefont {J.}~\bibnamefont
  {{Nunez-Iglesias}}}, \bibinfo {author} {\bibfnamefont {F.}~\bibnamefont
  {{B}oulogne}}, \bibinfo {author} {\bibfnamefont {J.~D.}\ \bibnamefont
  {{W}arner}}, \bibinfo {author} {\bibfnamefont {N.}~\bibnamefont {{Y}ager}},
  \bibinfo {author} {\bibfnamefont {E.}~\bibnamefont {{G}ouillart}}, \bibinfo
  {author} {\bibfnamefont {T.}~\bibnamefont {{Y}u}}, \ and\ \bibinfo {author}
  {\bibnamefont {the scikit-image contributors}},\ }\href {\doibase 10.7717/peerj.453} 
  {\bibfield  {journal} {\bibinfo  {journal} {PeerJ}\
  }\textbf {\bibinfo {volume} {2}},\ \bibinfo {pages} {e453} (\bibinfo {year}
  {2014})}\BibitemShut {NoStop}%
\end{thebibliography}
\end{document}